\documentclass{mn}
\newif\ifAMStwofonts
\usepackage{graphicx}

\def\gsim{\;\lower4pt\hbox{${\buildrel\displaystyle >\over\sim}$}\;}
\def\lsim{\;\lower4pt\hbox{${\buildrel\displaystyle <\over\sim}$}\;}
\def\grls{\;\lower4pt\hbox{${\buildrel\displaystyle >\over <}$}\;}

\ifoldfss
  \ifCUPmtlplainloaded \else
    \NewTextAlphabet{textbfit} {cmbxti10} {}
    \NewTextAlphabet{textbfss} {cmssbx10} {}
    \NewMathAlphabet{mathbfit} {cmbxti10} {} 
    \NewMathAlphabet{mathbfss} {cmssbx10} {} 
  \fi
  \ifAMStwofonts
    \ifCUPmtlplainloaded \else
      \NewSymbolFont{upmath} {eurm10}
      \NewSymbolFont{AMSa} {msam10}
      \NewMathSymbol{\upi}     {0}{upmath}{19}
      \NewMathSymbol{\umu}     {0}{upmath}{16}
      \NewMathSymbol{\upartial}{0}{upmath}{40}
      \NewMathSymbol{\leqslant}{3}{AMSa}{36}
      \NewMathSymbol{\geqslant}{3}{AMSa}{3E}

       \let\le=\leqslant
      \let\geq=\geqslant \let\ge=\geqslant
    \fi
  \fi
\fi 

\ifnfssone
  \newmathalphabet{\mathit}
  \addtoversion{normal}{\mathit}{cmr}{m}{it}
  \addtoversion{bold}{\mathit}{cmr}{bx}{it}
  \newmathalphabet{\mathbfit} 
  \addtoversion{normal}{\mathbfit}{cmr}{bx}{it}
  \addtoversion{bold}{\mathbfit}{cmr}{bx}{it}
  \newmathalphabet{\mathbfss} 
  \addtoversion{normal}{\mathbfss}{cmss}{bx}{n}
  \addtoversion{bold}{\mathbfss}{cmss}{bx}{n}
  \ifAMStwofonts
    \ifCUPmtlplainloaded \else
      \UseAMStwoboldmath
      \makeatletter
      \new@mathgroup\upmath@group
      \define@mathgroup\mv@normal\upmath@group{eur}{m}{n}
      \define@mathgroup\mv@bold\upmath@group{eur}{b}{n}
      \edef\UPM{\hexnumber\upmath@group}
      \new@mathgroup\amsa@group
      \define@mathgroup\mv@normal\amsa@group{msa}{m}{n}
      \define@mathgroup\mv@bold\amsa@group{msa}{m}{n}
      \edef\AMSa{\hexnumber\amsa@group}
      \makeatother
      \mathchardef\upi="0\UPM19
      \mathchardef\umu="0\UPM16
      \mathchardef\upartial="0\UPM40
      \mathchardef\leqslant="3\AMSa36
      \mathchardef\geqslant="3\AMSa3E

       \let\le=\leqslant
      \let\geq=\geqslant \let\ge=\geqslant
    \fi
  \fi
\fi 

\ifnfsstwo
  \DeclareMathAlphabet{\mathbfit}{OT1}{cmr}{bx}{it}
  \SetMathAlphabet\mathbfit{bold}{OT1}{cmr}{bx}{it}
  \DeclareMathAlphabet{\mathbfss}{OT1}{cmss}{bx}{n}
  \SetMathAlphabet\mathbfss{bold}{OT1}{cmss}{bx}{n}
  \ifAMStwofonts
    \ifCUPmtlplainloaded \else
      \DeclareSymbolFont{UPM}{U}{eur}{m}{n}
      \SetSymbolFont{UPM}{bold}{U}{eur}{b}{n}
      \DeclareSymbolFont{AMSa}{U}{msa}{m}{n}
      \DeclareMathSymbol{\upi}{0}{UPM}{"19}
      \DeclareMathSymbol{\umu}{0}{UPM}{"16}
      \DeclareMathSymbol{\upartial}{0}{UPM}{"40}
      \DeclareMathSymbol{\leqslant}{3}{AMSa}{"36}
      \DeclareMathSymbol{\geqslant}{3}{AMSa}{"3E}

       \let\le=\leqslant
      \let\geq=\geqslant \let\ge=\geqslant
    \fi
  \fi
\fi 

\ifCUPmtlplainloaded \else
  \ifAMStwofonts \else 
    \def\upi{\pi}
    \def\umu{\mu}
    \def\upartial{\partial}
  \fi
\fi

\title[Stationary MSID configurations]
{Stationary Nonaxisymmetric Configurations of Magnetized Singular
Isothermal Disks}

\author[Lou]
{
Yu-Qing Lou$^{1}$\\
$^1$National Astronomical Observatories, Chinese Academy of
Sciences, A20, Datun Road,
Beijing, 100012 China;\\
Department of Astronomy and Astrophysics, The University of
Chicago,
Chicago, Illinois 60637 USA;\\
Email: lou@oddjob.uchicago.edu; and\\
Physics Department, The Tsinghua Astrophysics Center,
           Tsinghua University, Beijing 100084 China.}

\date{Accepted ....
      Received ...;
      in original form ...}
\pagerange{\pageref{firstpage}--\pageref{lastpage}}

\pubyear{2001}

\begin{document}

\maketitle

\label{firstpage}

\begin{abstract}
We construct both aligned and unaligned (logarithmic spiral)
stationary configurations of nonaxisymmetric magnetohydrodynamic
(MHD) disks from either a full or a partial razor-thin power-law
axisymmetric magnetized singular isothermal disk (MSID) that is
embedded with a coplanar azimuthal magnetic field $B_{\theta}$ of
a non-force-free radial scaling $r^{-1/2}$ and that rotates
differentially with a flat rotation curve of speed $aD$, where $a$
is the isothermal sound speed and $D$ is the dimensionless
rotation parameter. Analytical solutions and stability criteria
for determining $D^2$ are derived. For aligned nonaxisymmetric
MSIDs, eccentric $m=1$ displacements may occur at arbitrary $D^2$
in a full MSID but are allowed only with $a^2D^2=C_A^2/2$ in a
partial MSID ($C_A$ is the Alfv\'en speed), while each case of
$|m|\ge 1$ gives two possible values of $D^2$ for purely azimuthal
propagations of fast and slow MHD density waves (FMDWs and SMDWs)
that appear stationary in an inertial frame of reference. For disk
galaxies modeled by a partial MSID resulting from a massive
dark-matter halo with a flat rotation curve and $a^2D^2\gg C_A^2$,
stationary aligned perturbations of $m=1$ are not allowed. For
unaligned logarithmic spiral MSIDs with $|m|\ge 1$, there exist
again two values of $D^2$, corresponding to FMDWs and SMDWs that
propagate in both radial and azimuthal directions relative to the
MSID and that appear stationary in an inertial frame of reference.
The larger $D^2$ is always physically valid, while the smaller
$D^2$ is valid only for $a>C_A/2$ with a positive surface mass
density $\Sigma_{0}$. For observational diagnostics, we examine
the spatial phase relationships among enhancements of gas density
and magnetic field as well as velocity perturbations. These
results are useful for probing magnetized bars, or lopsided,
normal, and barred spiral galaxies as well as for testing
numerical MHD codes. In the case of NGC 6946, interlaced optical
and magnetic field spiral patterns of SMDWs can persist in a disk
of flat rotation curve. Theoretical issues regarding the modal
formalism and the MSID perspective are also discussed.
\end{abstract}


\section{Introduction}

It is a challenge to study the large-scale dynamics of various
morphologies of disk galaxies such as bars and lopsided, barred,
and normal spiral structures (e.g., Baldwin, Lynden-Bell, \&
Sancisi 1980; Richter \& Sancisi 1994; Rix \& Zaritsky 1995). To
include gravitational interactions and magnetohydrodynamics (MHD)
of the interstellar medium (ISM) and magnetic field, we risk in
making the task even more formidable. However, multi-wavelength
observational diagnostics involving both ISM and magnetic field do
provide indispensable clues to the overall dynamics. For example,
there have been growing numbers of high-quality observations on
large-scale magnetic field structures in nearby spiral galaxies
(e.g., Sofue et al. 1986; Kronberg 1994; Beck et al. 1996; Beck
2001 and references therein). We here venture to formulate a
limited yet nontrivial theoretical MHD disk problem in which
%
stationary nonaxisymmetric MHD perturbation configurations are
constructed from a background axisymmetric MSID of interstellar
gas medium embedded with a coplanar azimuthal magnetic field. We
search for both aligned and unaligned stationary configurations in
a two-dimensional self-gravitating MSID with a flat rotation
curve.

Our model analysis here will give various stationary morphologies
in a magnetized gas disk, including bars and lopsided, barred, and
normal spiral structures. Moreover, we provide phase relationships
of spatial patterns among magnetic field, gas density, and
velocity perturbations that can be examined observationally.
Regarding the recent wavelet analysis on multi-wavelength data of
the spiral galaxy NGC 6946 (Frick et al. 2000, 2001) that revealed
an extension of the interlaced magnetic and optical spiral
structures into the outer disk with a largely flat rotation curve,
our analysis here conveys an important message that stationary
logarithmic spiral patterns of slow MHD density waves (SMDWs; Fan
\& Lou 1996; Lou \& Fan 1998a), with interlaced spiral
enhancements of magnetic field and gas density, can indeed persist
in an extended MSID with a flat rotation curve (Lou \& Fan 2002).

Conceptually, the stationary pattern problem here is closely tied
to the density wave problem in a differentially rotating disk
(Syer \& Tremaine 1996; Shu et al. 2000). It is the balance
between relevant wave pattern speeds and disk rotation that leads
to possible nonaxisymmetric patterns stationary in an inertial
frame of reference. Historically, the seminal idea of MHD density
waves
was contemporary (Lynden-Bell 1966; Roberts \& Yuan 1970) with the
early development of density wave theory four decades ago (Lin \&
Shu 1964, 1966; Goldreich \& Lynden-Bell 1965; Toomre 1969, 1977;
Lin 1967, 1987; Shu 1970a, b).
While the stellar disk provides a massive``template", MHD density
wave processes in the ISM disk (Fan \& Lou 1996, 1997, 1999; Lou
\& Fan 1997, 1998a, b, 2000a, b, 2002; Lou, Han, \& Fan 1999; LYF
2001; LYFL 2001) do have additional dynamic freedoms.

Theoretically, there exist complemetary perspectives, different
motivations, and independent approaches to study various bar
phenomena in rotating self-gravitating fluid bodies. The
masterpiece of Chandrasekhar (1969) on ellipsoidal figures of
equilibrium
summarizes the beautiful mathematical descriptions of Maclaurin,
Jacobi, Dedekind, Riemann ellipsoids and Poincar\'e pear-shaped
configurations as well as their close interrelations on the basis
of instability analyses and characteristics of bifurcations in
incompressible self-gravitating fluids. Complementary numerical
studies of compressible, self-gravitating, differentially
rotating, nonspherical equilibrium figures have shown striking
family resemblances to these classical exactly solved problems
(Ostriker 1978). By collapsing the dimension along the rotation
axis, one can study equilibria and stability properties of
two-dimensional Riemann disks of uniform rotation relevant to
central parts of disk galaxies (Weinberg \& Tremaine 1983;
Weinberg 1983). Another broad class of disk problems involves
stability properties of singular isothermal disks (SIDs) (Zang
1976; Toomre 1977; Lemos et al. 1991; Lynden-Bell \& Lemos 1993;
Goodman \& Evans 1999). Syer \& Tremaine (1996) found solutions to
a class of stationary nonaxisymmetric perturbation SID
configurations. Shu et al.
(2000) derived solutions for stationary perturbation
configurations in isopedically magnetized SIDs
and interpreted them as onsets of bar-type and barred-spiral
instabilities (Galli et al. 2001). Our analysis here parallels
that of Shu et al. (2000) but with a coplanar magnetic field in an
MSID, and we examine the MSID problem from the perspective of
stationary fast and slow MHD density waves (FMDWs and SMDWs; Fan
\& Lou 1996; Lou \& Fan 1998a; LYF 2001). We derive a form of
magnetic virial theorem for an MSID and suggest the ratio of
rotation energy to the sum of gravitational and magnetic energies
to be crucial for the MSID stability.

Shu et al. (2000) pursued the analogies of bar-type instabilities
known in thin self-gravitating disks (Hohl 1971; Miller et al.
1970) as well as in rotating self-gravitating spheroids and
ellipsoids of uniform density (Chandrasekhar 1969; Ostriker 1978)
for equilibria and instabilities of isopedically magnetized SIDs.
In spite of the known difficulties (Galli et al. 2001), the
fascinating scenario of the so-called ``fission theory" by
Poincar\'e (1885), Liapunov (1905), and Jeans (1928) that regards
incompressible ellipsoidal figures of equilibrium as potential
candidates for fissioning through pear-shaped equilibria into
binary stars has lead Shu et al. (2000) to speculate that
nonaxisymmetric, magnetized, compressible, and perhaps truncated
SIDs might prove to be promising candidates for fragmentation into
binary- and multiple-star systems. They also hypothesized that a
rapid loss of magnetic flux at a certain stage might hold the
crucial key of resolving or bypassing the relevant known
difficulties.
In line with these speculations, the present MSID problem should
be relevant to star formation research as well.

Based on theoretical analyses (Lau \& Bertin 1978; Lin \& Lau
1979; Bertin \& Mark 1979) to retain higher-order effects for the
WKBJ
expansion of the Poisson equation,
Bertin et al. (1989a, b) heuristically argued and invoked the
cubic dispersion relation of density waves in a thin fluid stellar
disk as the conceptual
basis for classifying barred and normal spiral galaxies (viz., the
Hubble classification scheme of galaxies; e.g., Lin \& Lau 1979)
constructed numerically by solving the standard linear
integro-differential density wave equations with chosen boundary
conditions.
Important aspects of this modal perspective have been
comprehensively summarized in Bertin \& Lin (1996). In their
scenario, bars and barred spirals are essentially viewed as
density-wave phenomena that are better described when the effects
of long-range self-gravity and differential rotation are more
fully included. Specifically, the cubic dispersion relation
becomes cubic in the radial wavenumber $k$ by including the
tangential shear force (TSF) for nonaxisymmetric coplanar
perturbations (LYF 2001). Besides the familiar short- and
long-wave branches (both are somewhat modified by the TSF), there
is now a third wave branch of ``open modes" characterized by even
smaller $k$ that bear striking resemblances to bars and barred
spirals when superposed with an axisymmetric bulge (see Fig. 14 of
Bertin et al. 1989a; Lin 1996 private communications). As we will
solve the Poisson integral exactly, it would be natural to pursue
the correspondence between the modal perspective of Bertin \& Lin
(1996) and the (M)SID results here as well as of Shu et al.

This paper is structured as follows. We formulate in \S 2 the
problems of full and partial MSIDs and display relevant MHD
equations. Solutions and analyses for aligned and unaligned MSID
configurations are presented in \S 3. In \S 4, we examine the
spatial phase relationships between enhancements of gas density,
magnetic field, and velocity disturbances. In \S 5, we discuss the
connection between the modal formalism and the MSID perspective
for classifying normal and barred spiral galaxies, indicate the
implication to multi-band observations of spiral galaxy NGC 6946,
and summarize the results. Mathematical formulae are collected in
Appendices A$-$E for the convenience of reference.

\section{Formulation of the MSID Problem}

The key difference of our formulation and that of Shu et al.
(2000) is that their magnetic field is poloidal threading across
the disk and may be effectively relegated into two dimensionless
parameters\footnote{Specifically, $\Theta\equiv
(\lambda^2+1+2\eta^2)/(\lambda^2+\eta^2)$ and $\epsilon\equiv
1-\lambda^{-2}$ with $\lambda\equiv 2\pi G^{1/2}\Sigma/B_z=$
const. and $\eta\equiv |\vec g_{\parallel}|/(2\pi G\Sigma)$ where
$|\vec g_{\parallel}|$ is the coplanar gravitational
acceleration.} $\Theta$ and $\epsilon$ (Schmitz 1987; Shu \& Li
1997) such that $a^2\rightarrow\Theta a^2$ and
$G\rightarrow\epsilon G$ ($a$ is the isothermal sound speed and
$G$ is the gravitational constant), whereas the non-force-free
magnetic field of our model is azimuthal and coplanar with the
disk. In essence, the effect of their poloidal magnetic field can
be scaled away such that the analysis is equivalent to a
hydrodynamic one, while the coplanar magnetic field of our model
can give rise to new classes or features of stationary
nonaxisymmetric MSID configurations. We refer to our model as an
MSID problem, although the SID problem as formulated by Shu et al.
(2000) is isopedically magnetized, and proceed to examine several
specific aspects in an orderly manner.

\subsection{Nonlinear unsteady MHD equations}

For large-scale MHD processes, dissipative effects may be ignored
as a first approximation. In cylindrical coordinates
$(r,\theta,z)$, ideal time-dependent MHD equations include
$$
{\partial\Sigma\over\partial t} +{1\over r}{\partial (r\Sigma
u)\over\partial r} +{1\over r^2}{\partial (\Sigma j)\over
\partial\theta}=0\ \eqno(2.1.1)
$$
for the mass conservation, where $\Sigma$ is the vertically
integrated mass density, $u$ is the radial velocity, $j\equiv rv$
is the specific angular momentum in the vertical $\hat z$
direction and $v$ is the azimuthal velocity in
$\hat\theta-$direction,
\begin{eqnarray}
& & {\partial u\over \partial t} +u{\partial u\over\partial r}
+{j\over r^2}{\partial u\over\partial\theta} -{j^2\over r^3}
=-{1\over\Sigma}{\partial\Pi\over\partial r} \nonumber \\ &
&\qquad\quad -{\partial\phi_T\over\partial r} -{1\over\Sigma}\int
{dzB_{\theta}\over 4\pi r} \bigg[{\partial
(rB_{\theta})\over\partial r} -{\partial
B_r\over\partial\theta}\bigg]\quad\ (2.1.2) \nonumber
\end{eqnarray}
for the radial momentum equation, where $\Pi$ is the vertically
integrated gas pressure and $\phi_T$ is the total gravitational
potential inclusive of that from a possible dark-matter halo,
\begin{eqnarray}
& & {\partial j\over\partial t} +u{\partial j\over\partial r}
+{j\over r^2}{\partial j\over\partial\theta}
=-{1\over\Sigma}{\partial \Pi\over\partial\theta} \nonumber \\ & &
\quad\qquad -{\partial\phi_T\over\partial\theta}
+{1\over\Sigma}\int {dz B_r\over 4\pi} \bigg[{\partial
(rB_{\theta})\over\partial r} -{\partial
B_r\over\partial\theta}\bigg]\quad\ (2.1.3) \nonumber
\end{eqnarray}
for the azimuthal momentum equation in $\hat\theta-$direction,
$$
F\phi_T
=-G\oint d\psi\!\int_{0}^{\infty}\!\! {\Sigma(\zeta,\psi, t)\zeta
d\zeta \over [\zeta^2+r^2-2\zeta
r\cos(\psi-\theta)]^{1/2}}\eqno(2.1.4)
$$
for the Poisson integral equation where $F\phi_T$ is the
gravitational potential from the gas disk with $0\le F\le 1\ $,
$$
{\partial (rB_r)\over\partial r} +{\partial
B_{\theta}\over\partial\theta}=0\eqno(2.1.5)
$$
for the divergence-free condition on $\vec B\equiv
(B_r,B_{\theta},0)$,
$$
{\partial B_r\over\partial t} ={1\over
r}{\partial\over\partial\theta} (uB_{\theta}-vB_r)\eqno(2.1.6)
$$
for the radial magnetic field induction equation, and
$$
{\partial B_{\theta}\over\partial t} =-{\partial\over\partial
r}(uB_{\theta}-vB_r)\eqno(2.1.7)
$$
for the azimuthal magnetic field induction equation.
The polytropic approximation $\Pi=a^2\Sigma$ is invoked as usual
where $a$ is the isothermal sound speed.

For a static vertical balance without a vertical flow velocity
$v_z$, the vertical force balance
takes the form of
$$
0=-\int dz {\partial p\over\partial z} -\int
dz\rho{\partial\phi\over\partial z} -\int dz
{\partial\over\partial z} {(B_{\theta}^2+B_r^2)\over 8\pi}\
,\eqno(2.1.8)
$$
where gas and magnetic pressures work together against the
vertical gravity towards the disk at $z=0$.

\subsection{Equilibrium of an axisymmetric MSID}

For an MSID geometry of axisymmetry, one solves the MHD equations
with the Poisson integral (2.1.4).
We presume a flat rotation curve with a background disk angular
rotation speed $ \Omega=aD/r\ ,
$ where $D$ is the dimensionless rotation parameter. The cases of
$D>1$ and $D<1$ correspond then to supersonic and subsonic MSID
rotations. To avoid the magnetic field winding dilemma (Lou \& Fan
1998a), the background magnetic field is taken to be azimuthal
about the symmetry axis with a scaling of $ B_{\theta}={\cal
F}r^{-1/2}
$ where ${\cal F}$ is a constant. The Lorentz force of this
$B_{\theta}$ profile \footnote{$B_{\theta}$ scales as $r^{-1}$ for
a force-free azimuthal magnetic field.} is included in the MSID
radial force balance. The vertically integrated gas pressure
$\Pi_{0}$ of the background
is related to the background surface mass density $\Sigma_0$ by $
\Pi_{0}=a^2\Sigma_{0}.
$ The epicyclic oscillation frequency $\kappa$ of a flat rotation
curve is defined by $ \kappa^2\equiv
(2\Omega/r)d(r^2\Omega)/dr=2\Omega^2 ,
$ and the Alfv\'en speed $C_A$ in an MSID is defined by $
C_A^2\equiv \int dz B_{\theta}^2/(4\pi\Sigma_{0}).
$

%
%

The rotation speed of a disk galaxy $V_{\theta}\equiv \Omega
r=aD\gg C_A$ due to the presence of a massive dark-matter halo
(presumed to be axisymmetric here). By the equipartition argument,
the thermal and magnetic energy densities are comparable with
$a\sim C_A$. It then follows that $D\gg 1$ for supersonic and
super-Alfv\'enic rotations in typical disk galaxies.

To attribute a fraction $(1-F)$ of the total gravity in the
background equilibrium to an axisymmetric dark-matter halo that is
unresponsive to gas disk perturbations, one may write $
\partial\phi_T/\partial r=F\partial\phi_T/\partial r
+(1-F)\partial\phi_T/\partial r\ .
$
The case of $F=1$ is referred to as a full (M)SID, while the case
of $0\le F<1$ is referred to as a partial (M)SID (Syer \& Tremaine
1996; Shu et al. 2000). In reference to equations (5), (6), (21)
and (27) of Syer \& Tremaine (1996), their parameter $f$ is
related to $F$ here by $f\equiv (1-F)/F$. The background
rotational equilibrium of an MSID requires
$$
-{a^2D^2\over r}={a^2\over r} -{\partial\phi_T\over\partial
r}-{C_A^2\over 2r}\ ,\eqno(2.2.1)
$$
where the portion of the gravitational potential associated with
the gas disk satisfies $ F\partial\phi_T/\partial r=2\pi
G\Sigma_{0}
$ in the disk plane at $z=0$, as required by the Poisson equation
in a razor-thin MSID. In force balance (2.2.1), the gravity and
net Lorentz forces are radially inward (i.e., an outward magnetic
pressure force but a stronger inward magnetic tension force),
while the gas pressure and centrifugal forces are radially
outward. It follows for an MSID that
$$
\Sigma_{0}=F[a^2(1+D^2)-C_A^2/2]/(2\pi Gr)\ .\eqno(2.2.2)
$$
%
%
In the context of a protostellar disk, one cannot invoke a
dark-matter halo and one thus deals with a full (M)SID of $F=1$.
Shu et al. (2000; also Galli et al. 2001) recently performed
extensive analysis on constructions of stationary nonaxisymmetric
perturbation solutions from a background axisymmetric SID that is
isopedically magnetized with a poloidal magnetic field. These
solutions can be classified as aligned and unaligned
configurations (Kalnajs 1973) that require specific values of
$D^2$ for their very existence. These results in idealized
theoretical settings are interesting and important in that these
solutions, with proper interpretations, may be pertinent to
configurations of bars, barred spirals, and normal spirals in a
differentially rotating disk, and may bear implications to the
onset of bar or bar-type instabilities that lead to formation of
these possible configurations. For example, for the aligned case
of nonaxisymmetric stationary perturbations (of angular variation
$\exp[-im\theta]$ with an integer $m$) studied by Shu et al.
(2000) (see their eqns $[25]-[27]$), one finds an unconstrained or
arbitrary $D^2$ for $|m|=1$ (see also equation [46] of Syer \&
Tremaine 1996), and $ D^2=|m|/(|m|+2)\
$ for $|m|\ge 2$. The latter with $D^2<1$
implies a subsonic protostellar disk rotation.

\subsection{Coplanar MHD perturbations in an MSID}

From equations $(2.1.1)-(2.1.7)$, one derives coplanar MHD
perturbation equations in an MSID of gas,
$$
{\partial\Sigma_1\over\partial t} +{1\over r}{\partial
(r\Sigma_{0} u_1)\over\partial r}
+\Omega{\partial\Sigma_1\over\partial\theta} +{\Sigma_{0}\over
r^2}{\partial j_1\over \partial\theta}=0\ , \eqno(2.3.1)
$$
\begin{eqnarray}
& & {\partial u_1\over\partial t} +\Omega {\partial
u_1\over\partial\theta} -{2\Omega j_1\over
r}=-{\partial\over\partial r}
\bigg({a^2\Sigma_1\over\Sigma_{0}}+\phi_1\bigg) \nonumber \\ &
&\qquad +{C_A^2\Sigma_1\over 2\Sigma_{0}r} -{1\over\Sigma_{0}}\int
{dzB_{\theta}\over 4\pi r} \bigg[{\partial
(rb_{\theta})\over\partial r} -{\partial
b_r\over\partial\theta}\bigg] \nonumber \\ & &\qquad\qquad\qquad
-{1\over\Sigma_{0}}\int {dz b_{\theta}\over 4\pi r} {\partial
(rB_{\theta})\over\partial r}\ , \qquad\qquad\quad
(2.3.2)\nonumber
\end{eqnarray}
\begin{eqnarray}
& & {\partial j_1\over\partial t} +{r\kappa^2\over 2\Omega}u_1
+\Omega {\partial j_1\over\partial\theta}
=-{\partial\over\partial\theta}
\bigg({a^2\Sigma_1\over\Sigma_{0}}+\phi_1\bigg) \nonumber \\ &
&\qquad\qquad\qquad +{1\over \Sigma_{0}}\int {dz b_r\over 4\pi}
{\partial (rB_{\theta})\over\partial r}\ , \qquad\qquad\quad
(2.3.3)\nonumber
\end{eqnarray}
$$
\phi_1
=-G\oint d\psi\!\int_{0}^{\infty}\!\! {\Sigma_1(\zeta,\psi,
t)\zeta d\zeta \over [\zeta^2+r^2-2\zeta
r\cos(\psi-\theta)]^{1/2}}\ , \eqno(2.3.4)
$$
$$
{\partial (rb_r)\over\partial r} +{\partial
b_{\theta}\over\partial\theta}=0\ ,\eqno(2.3.5)
$$
$$
{\partial b_r\over\partial t} ={1\over
r}{\partial\over\partial\theta} (u_1B_{\theta}-r\Omega b_r)\
,\eqno(2.3.6)
$$
$$
{\partial b_{\theta}\over\partial t} =-{\partial\over\partial r}
(u_1B_{\theta}-r\Omega b_r)\ ,\eqno(2.3.7)
$$
where $\vec b\equiv (b_r,b_{\theta},0)$ is the coplanar magnetic
field perturbation, $\phi_1$ is the perturbation of $F\phi_T$ from
gas distribution, and other variables with subscript $_1$ are
perturbations to the pertinent equilibrium variables. Here, we set
$v_z=b_z=0$ and do not consider vertical variations across the
disk. All with the harmonic $\exp(i\omega t-im\theta)$ dependence,
we introduce complex radial variations $S(r)$, $U(r)$, $J(r)$,
$V(r)$, $R(r)$, and $Z(r)$ for $\Sigma_1$, $u_1$, $j_1$, $\phi_1$,
$b_r$, and $b_z$.
Coplanar MHD perturbation equations $(2.3.1)-(2.3.7)$ can then be
reduced to
$$
i(\omega-m\Omega)S+{1\over r} {\partial\over\partial
r}(r\Sigma_{0}U) -{im\Sigma_{0}\over r^2}J=0\ ,\eqno(2.3.8)
$$
\begin{eqnarray}
& & \!\!\!\!\!\!\!\!\!\!\! i(\omega -m\Omega)U-{2\Omega J\over r}
=-{\partial\Phi\over\partial r} -{1\over\Sigma_{0}}\int {dz Z\over
4\pi r} {\partial (rB_{\theta})\over\partial r} \nonumber \\ &
&\quad +{C_A^2S\over 2\Sigma_{0}r}-{1\over\Sigma_{0}} \int {dz
B_{\theta}\over 4\pi r}\bigg[ {\partial (rZ)\over\partial
r}+imR\bigg]\ , \ \qquad (2.3.9)\nonumber
\end{eqnarray}
where
$ \Phi\equiv a^2S/\Sigma_{0}+V\ ,
$
$$
i(\omega-m\Omega)J+{r\kappa^2\over 2\Omega}U
=im\Phi+{1\over\Sigma_{0}}\int {dz R\over 4\pi} {\partial
(rB_{\theta})\over\partial r}\ ,\eqno(2.3.10)
$$
$$
V(r)=-G\oint d\chi\int_{0}^{\infty} {S(\zeta)\cos(m\chi)\zeta
d\zeta\over [\zeta^2+r^2-2\zeta r\cos\chi]^{1/2}}\ ,\eqno(2.3.11)
$$
$$
{\partial (rR)\over\partial r}-imZ=0\ ,\eqno(2.3.12)
$$
$$
i(\omega-m\Omega)R+{imB_{\theta}\over r}U=0\ ,\eqno(2.3.13)
$$
$$
i\omega Z={\partial\over\partial r}(r\Omega R)
-{\partial\over\partial r}(B_{\theta}U)\ .\eqno(2.3.14)
$$
It suffices to use only two of the three equations
$(2.3.12)-(2.3.14)$ for magnetic field perturbation $\vec
b=(b_r,b_{\theta},0)$.


Rearrangement of equations $(2.3.8)-(2.3.14)$ in terms of $S$,
$V$, $J$ and $iU$
further leads to
$$
Z=-{i\over m}{\partial (rR)\over\partial r}\ \eqno(2.3.15)
$$
from the divergence-free condition (2.3.12) of $\vec b$, and
$$
R=-{mB_{\theta}U\over r(\omega-m\Omega)}\ \eqno(2.3.16)
$$
from the radial magnetic induction equation (2.3.13). Using
equations (2.3.15) and (2.3.16) in the radial and azimuthal
momentum equations (2.3.9) and (2.3.10), one derives
\begin{eqnarray}
& & i(\omega-m\Omega)U-{2\Omega J\over r}
=-{\partial\Phi\over\partial r} +{C_A^2S\over 2\Sigma_{0}r}
\nonumber \\ & &\qquad -iC_A^2 \Bigg[\!\!\!\Bigg[ {1\over
r^{1/2}}{\partial\over\partial r} \bigg\lbrace
r{\partial\over\partial r} \bigg[ {U\over
r^{1/2}(\omega-m\Omega)}\bigg]\bigg\rbrace \nonumber \\ & &
-{m^2U\over r^2(\omega-m\Omega)} +{1\over 2
r^{1/2}}{\partial\over\partial r} \bigg [{U\over
r^{1/2}(\omega-m\Omega)}\bigg] \Bigg]\!\!\!\Bigg]\ \hbox{ }\
(2.3.17)\nonumber
\end{eqnarray}
and
$$
i(\omega-m\Omega)J+{r\kappa^2\over 2\Omega}U =im\Phi
-{mC_A^2U\over 2r(\omega-m\Omega)}\ .\eqno(2.3.18)
$$

\subsection{Stationary MHD perturbations in an MSID}

For nonaxisymmetric MHD perturbations stationary in an inertial
frame of reference, we set $\omega=0$
in equations (2.3.8), $(2.3.16)-(2.3.18)$ and obtain
$$
m\Omega S+{1\over r} {\partial\over\partial r}(r\Sigma_{0}iU)
+{m\Sigma_{0}\over r^2}J=0\ ,\eqno(2.4.1)
$$
\begin{eqnarray}
& &\!\!\!\!\!\! m\Omega iU+{2\Omega J\over r}
={\partial\Phi\over\partial r} -{C_A^2S\over 2\Sigma_{0}r}
-{C_A^2\over 2 r^{1/2}}{\partial\over\partial r} \bigg ({iU\over
m\Omega r^{1/2}}\bigg) \nonumber \\ & &\quad -{C_A^2\over
r^{1/2}}{\partial\over\partial r} \bigg [r{\partial\over\partial
r} \bigg( {iU\over m\Omega r^{1/2}}\bigg)\bigg ]
+{C_A^2miU\over\Omega r^{2}}\ , \qquad\ \!\!  (2.4.2)\nonumber
\end{eqnarray}
$$
m\Omega J+{r\kappa^2\over 2\Omega}iU =-m\Phi +{C_A^2iU\over
2\Omega r}\ ,\eqno(2.4.3)
$$
$$
iR={B_{\theta}iU\over\Omega r}\ ,\eqno(2.4.4)
$$
together with equations (2.3.11), (2.3.15), and the definition of
$\Phi$ after equation (2.3.9). As the pattern speed
$\omega_p\equiv\omega/m$ of possible MHD density waves is set to
zero in an inertial frame of reference {\it a priori}, these
equations are to be solved to determine the proper values of the
dimensionless rotation parameter $D$. As the system may support
possible FMDWs and SMDWs (Fan \& Lou 1996; Lou \& Fan 1998a), two
proper values of $D$ are expected.

\section{Solution Analysis of
Stationary MSID Configurations }

\subsection{The case of aligned perturbations}

The case of $m=0$ should be handled with care. It would be
misleading to use equations (2.4.3) and (2.4.4). One should
examine equations (2.3.8)$-$(2.3.14) with $\omega=m=0$. By setting
$U=R=0$, equations (2.3.8), (2.3.10), (2.3.12) and (2.3.14) are
satisfied, and equation (2.3.13) is identically zero. There is no
constraint on $Z$.
However, by setting $Z\propto r^{-1/2}$, $S\propto r^{-1}$,
$J\propto r$, $V\propto\ln r$, and $\Phi\propto \hbox{ const.}+\ln
r$, remaining equations (2.3.9) and (2.3.11) can be made
consistent with a rescaling of the axisymmetric background MSID.

For the nonaxisymmetric aligned case with $m\neq 0$, one takes the
density-potential pair
$$
rS=\hbox{const.}\ \eqno(3.1.1)
$$
and
$$
V=-(2\pi G/|m|)rS=\hbox{const.}\ \eqno(3.1.2)
$$
(e.g., Shu et al. 2000; Galli et al. 2001). For a constant $iU$
(see eq. [3.1.4] below), the mass conservation $(2.4.1)$ requires
$$
J=-\Omega r^2S/\Sigma_{0}\ ,\eqno(3.1.3)
$$
and the azimuthal momentum equation (2.4.3) gives consistently a
constant $iU$ by
$$
iU=m(\Phi-\Omega^2r^2S/\Sigma_{0}) /[C_A^2/(2\Omega r)-\Omega r]\
.\eqno(3.1.4)
$$
By equations (3.1.1), (3.1.2), and (2.2.2), it follows that $
\Phi\equiv a^2S/\Sigma_{0}+V=\hbox{const.}\
$ and a combination of equations (3.1.3) and (2.4.2) leads to
$$
\bigg[m\Omega r-{C_A^2(m^2-1/2)\over m\Omega r}\bigg]iU
-{2\Omega^2r^2S\over\Sigma_{0}}+{C_A^2S\over 2\Sigma_{0}}=0\ .
\eqno(3.1.5)
$$
Substitutions of expressions (3.1.2)$-$(3.1.4) into equation
(3.1.5) give the {\it solution condition} or {\it stationary
dispersion relation} for aligned nonaxisymmetric MHD
perturbations,
\begin{eqnarray}
& &\!\!\!\!\!\!\!\!\! {m^2\Omega r-C_A^2(m^2-1/2)/(\Omega r) \over
C^2_A/(2\Omega r)-\Omega r} \bigg({a^2\over\Sigma_{0}r}-{2\pi
G\over|m|} -{\Omega^2r\over\Sigma_{0}}\bigg) \nonumber \\ &
&\qquad\qquad\qquad -{2\Omega^2r\over\Sigma_{0}} +{C_A^2\over
2\Sigma_{0}r}=0\ .\qquad\qquad\quad\ \ (3.1.6) \nonumber
\end{eqnarray}
For a later examination of the spatial phase relationship between
$b_{\theta}$ and $\Sigma_1$, we further derive the following
results. By equations (2.3.15) and (2.4.4), the $\theta-$component
of the magnetic field perturbation is related to $iU$ by
$$
Z=-iUB_{\theta}/(2m\Omega r)\ .\eqno(3.1.7)
$$
By using expression (3.1.4) for constant $iU$ and expression of
constant $\Phi$ in equation (3.1.7), one obtains
$$
Z=-{B_{\theta}S\over 2\Sigma_{0}} {(a^2-2\pi G\Sigma_{0}
r/|m|-\Omega^2r^2) \over (C_A^2/2-\Omega^2r^2)}\ .\eqno(3.1.8)
$$
The solution criterion (3.1.6) may then be written as
$$
{a^2-2\pi G\Sigma_{0} r/|m|-\Omega^2r^2 \over
C_A^2/2-\Omega^2r^2}={2\Omega^2r^2-C_A^2/2 \over
m^2\Omega^2r^2-C_A^2(m^2-1/2)}\ .\eqno(3.1.9)
$$
By equation (3.1.8), the sign of either left-hand side (LHS) or
right-hand side (RHS) of equation (3.1.9) will determine the phase
relation between the surface mass density perturbation $S$ and the
azimuthal magnetic field perturbation $Z$.

\subsection{The physical nature of the solution criterion}

To fully understand the nature of solution condition (3.1.6), we
examine two conceptually related cases in order. The first case of
$C_A^2=0$ is essentially the one studied by Shu et al. (2000)
($\epsilon=1$ and $\Theta=1$) and can be explained as purely
azimuthal propagation of hydrodynamic density waves (Lin \& Shu
1964, 1966). By this clue, we proceed to show that the second case
of $C_A^2\neq 0$ corresponds to two possible situations of purely
azimuthal propagations of FMDWs and SMDWs (Lou \& Fan 1998a).

\subsubsection{The aligned full and partial SID cases}

We have verified with $C_A^2=0$ and $F=1$ in our analysis that the
results $(25)-(27)$ of Shu et al. (2000) come out naturally,
namely, either $D^2$ is arbitrary for $|m|=1$ or $D^2=|m|/(|m|+2)$
for $|m|\geq 2$. The latter with $D^2<1$ corresponds to a subsonic
SID rotation.

Moreover, the solution condition (3.1.6) or (3.1.9), with
$C_A^2=0$, can be written in the informative form of
$$
m^2\Omega^2=2\Omega^2+m^2a^2/r^2 -2\pi G\Sigma_{0}|m|/r\
.\eqno(3.2.1)
$$
In reference to the well-known WKBJ dispersion relation of density
waves (Lin \& Shu 1964, 1966 or equation [39] of Shu et al. 2000),
equation (3.2.1) can be readily obtained by replacing the radial
wavenumber $|k|$ with the azimuthal wavenumber $|m|/r$ and setting
$\omega=0$ in an inertial frame of reference. This clearly
describes an azimuthal propagation of hydrodynamic density waves,
and a stationary pattern in the sidereal frame of reference
requires specific values of $D^2$ for different $|m|$ values. It
was pointed out by Shu et al. (2000) that perturbations of the
aligned case has no radial wave propagation. We here provide a
transparent physical interpretation for the aligned case of
nonaxisymmetric stationary perturbations in terms of a purely
azimuthal density wave propagation (retrograde relative to the
disk) advected by the SID rotation such that the stationarity is
sustained. As the corotation is at infinity for a stationary
perturbation and $\kappa^2=2\Omega^2$, a solution of equation
(3.2.1) appears outside the Lindblad resonances also located at
infinity. Specifically, the solution is valid inside the inner
Lindblad resonance (ILR) in a finite radial range. Formally, only
the solution
of $|m|=1$ appears within the Lindblad resonances.

Equation (3.2.1) is quadratic in $|m|$ and the two possible values
of $|m|$ should be reminiscent of the long- and short-branches of
density waves in terms of the radial wavelength $2\pi/|k|$ (e.g.,
Binney \& Tremaine 1987), even though the $m^2\Omega^2$ term on
the LHS of equation (3.2.1) is now involved in determining the
proper values of $|m|$. By this perspective and for a full SID
with $F=1$, one might view the two values of $|m|$ given by
equation (3.2.1), namely $ |m|=1
$ with an arbitrary $D^2$ value, and $ |m|=2D^2/(1-D^2)
$ with $|m|\ge 2$, as ``long" and ``short" stationary azimuthally
propagating density waves. By the latter, one requires $D^2<1$ and
in the limit of $D^2\rightarrow 1$, the value of $|m|$ increases
to infinity corresponding to extremely short azimuthal
wavelengths. Perhaps, the $|m|=2$ case is the most interesting one
that mimics a stationary {\it barred configuration}.

For a partial SID with $0\le F<1$
and $C_A^2=0$, criterion (3.2.1) becomes
$$
m^2D^2=2D^2+m^2-|m|F(D^2+1)\ .\eqno(3.2.2)
$$
The two values of $|m|$ are then given by
$$
|m|={-F(D^2+1)\pm [F^2(D^2+1)^2+8D^2(D^2-1)]^{1/2} \over
2(D^2-1)}\ \eqno(3.2.3)
$$
where negative values of $|m|$ must be rejected.

To examine the solution property of (3.2.3) for $|m|$, condition
(3.2.2) may be cast into the revealing form of
$$
D^2=|m|(|m|-F)/(m^2+F|m|-2)\ .\eqno(3.2.4)
$$
For a full SID with $F=1$ in equation (3.2.4), one would have $
D^2=|m|/(|m|+2)<1
$ for a subsonic SID rotation. For $F=1$ and $|m|=1$ in equation
(3.2.4), the value of $D^2$ is arbitrary for a nontrivial
eccentric solution (Shu et al. 2000), including the case of a cold
rotating disk of $a^2=0$ but $\Omega r=aD\neq 0$ (see discussions
of Syer \& Tremaine 1996 about their eq. [43]) and the case of a
non-rotating disk of $D^2=0$ (see discussions of Syer \& Tremaine
1996 about their eq. [45]). In the limit of $F=0$, equation
(3.2.4) gives
$$
D^2=m^2/(m^2-2) > 1\ \eqno(3.2.5)
$$
(see eqns. [8], [21], [25], [44] of Syer \& Tremaine 1996 with
their $\beta=0$) for a supersonic SID rotation with $|m|\ge 2$,
and the case of $|m|=1$ is forbidden because $D^2>0$ is required.

Similarly for $0<F<1$, the case of $|m|=1$ is no longer allowed
for a partial SID because $D^2=-1$ is unphysical. In this context,
we note that the corollary of Syer \& Tremaine (1996) after their
equation (44) is inaccurate in the sense that the situation of
their $\beta=0$ (flat rotation curve) and finite $f\neq 0$
($0<F<1$) does not allow for aligned $m=1$ perturbation solutions.
There are two distinct classes of solutions for $|m|$ in general.
For $|m|<F^{-1}$, one would have a supersonic SID rotation with
$D^2>1$, while for $|m|>F^{-1}$, one would have a subsonic SID
rotation with $D^2<1$. It is then clear that the case of $F=1$
excludes the possibility of $D^2>1$, while the case of $F=0$
excludes the possibility of $D^2<1$. For $0\le F<1$, all solutions
with $|m|\ge 2$ satisfying condition (3.2.2) appears inside the
ILR.

For spiral galaxies, a partial SID model with $0\le F<1$ is more
relevant due to the presence of a massive dark-matter halo and
there seems to be no need to worry about the $|m|=1$ case of
aligned eccentric displacements because they only exist for a full
SID with $F=1$ (Zang 1976; Toomre 1977; Shu et al. 2000). However,
in star-forming clouds not involving an axisymmetric dark-matter
halo, such case of $|m|=1$ for aligned eccentric displacements
might lead to alternative SID configurations of interest (Shu et
al. 2000).

\subsubsection{Aligned MSID perturbation configurations}

With $C_A^2\neq 0$, the stationary criterion (3.1.6) or (3.1.9)
for the aligned case may be cast into the informative form of
\begin{eqnarray}
& & \!\!\!\!\!\!\!\!\!\!
m^4\Omega^4-\bigg[2\Omega^2+\bigg({C_A^2\over r^2} +{a^2\over
r^2}-{2\pi G\Sigma_{0}\over |m|r}\bigg)m^2 -{2C_A^2\over
r^2}\bigg]m^2\Omega^2 \nonumber \\ & &\!\!\!\!\!\!\!\!\!\!
+{m^2C_A^2\over r^2}\bigg[\bigg({a^2\over r^2} -{2\pi
G\Sigma_{0}\over |m|r}\bigg) \bigg(m^2-{1\over
2}\bigg)-{C_A^2\over 4r^2}\bigg]=0,\ (3.2.6)\nonumber
\end{eqnarray}
which reminds us of the dispersion relation for spiral FMDWs and
SMDWs (Fan \& Lou 1996; Lou \& Fan 1998a; LYF 2001) by simply
replacing the radial wavenumber $|k|$ with the azimuthal
wavenumber $|m|/r$ and by setting $\omega=0$ in an inertial frame
of reference. In the earlier derivation of WKBJ dispersion
relation for FMDWs and SMDWs, we took a force-free
$B_{\theta}\propto r^{-1}$. The present case of $B_{\theta}\propto
r^{-1/2}$ is not force-free and some extra terms in expression
(3.2.6) are due to the additional radial and azimuthal Lorentz
force terms as well as geometric effects in equations (2.3.2) and
(2.3.3). Physically, the two possible $D^2$ values contained in
equation (3.2.6) should relate to the two situations of purely
azimuthal propagation of FMDWs and SMDWs (retrograde relative to
the disk) such that the disk rotation renders them stationary in
an inertial frame of reference.


The equivalent criteria (3.1.6), (3.1.9), and (3.2.6) may be
arranged into the form of
\begin{eqnarray}
& &\!\!\!\!\!\!\! [m^2a^2D^2-C_A^2(m^2-1/2)] \nonumber \\ & &
\times [a^2(1-D^2)-F(a^2D^2+a^2-C_A^2/2)/|m|] \nonumber \\ &
&\qquad +(2a^2D^2-C_A^2/2)(a^2D^2-C_A^2/2)=0\ .\qquad
(3.2.7)\nonumber
\end{eqnarray}
For a full MSID with $F=1$, the LHS of expression (3.2.7) can be
factored into
\begin{eqnarray}
& & (|m|-1)\lbrace (a^2D^2-C_A^2)(1-D^2)a^2|m|^2 \nonumber \\ &
&\qquad +(a^2D^2-C_A^2)(C_A^2/2-2a^2D^2)|m| \nonumber \\ & &\qquad
-[C_A^2/2-a^2(D^2+1)]C_A^2/2\rbrace=0\ . \qquad\qquad (3.2.8)
\nonumber
\end{eqnarray}
Therefore, similar to the case of perturbations in an isopedically
magnetized full SID (Shu et al. 2000), the $D^2$ value is
unconstrained for $|m|=1$. For $|m|\ge 1$, the second factor on
the LHS of equation (3.2.8) may be rearranged into
\begin{eqnarray}
& &\!\!\!\!\!\!\!\!\!\!\!\!\!\!\!
|m|(2+|m|)a^4D^4-[C_A^2/2+5C_A^2|m|/2 +(a^2+C_A^2)|m|^2]a^2D^2
\nonumber \\ & &\!\!\!\!\!\!
+(C_A^2/2-a^2)C_A^2/2+C_A^4|m|/2+C_A^2a^2|m|^2=0\ \ \ \
(3.2.9)\nonumber
\end{eqnarray}
with the last coefficient being positive for $|m|\ge 1$.
By (3.2.9), there are two positive values of $a^2D^2$, namely
\begin{eqnarray}
& & a^2D^2={C_A^2/2+5C_A^2|m|/2+(a^2+C_A^2)|m|^2 \over
2|m|(2+|m|)} \nonumber \\ & &\qquad \pm\bigg\lbrace
\bigg[{C_A^2\over 2}+{5C_A^2|m|\over 2} +(a^2+C_A^2)|m|^2\bigg]^2
\nonumber \\ & &\qquad\qquad -4|m|(2+|m|)\bigg[{C_A^2\over 2}
\bigg({C_A^2\over 2}-a^2\bigg) \nonumber \\ & & +{C_A^4|m|\over 2}
+C_A^2a^2|m|^2\bigg]\bigg\rbrace^{1/2}[2|m|(2+|m|)]^{-1}, \ \
(3.2.10) \nonumber
\end{eqnarray}
where the determinant $\Delta$ is non-negative
for $|m|\geq 1$ (see Appendix A or equation [$D9$]).

For a partial MSID with $0\le F<1$, expression (3.2.7) may be cast
into the form of
\begin{eqnarray}
& &\!\!\!\!\!\!\!\!\! (2-m^2-F|m|)a^4D^4 +({\cal A}-{\cal B}+{\cal
C})a^2D^2
\nonumber \\ & &\qquad
-(C_A^2/m^2)(m^2-1/2){\cal A}+C_A^4/4=0\ , \qquad (3.2.11)
\nonumber
\end{eqnarray}
where ${\cal A}$, ${\cal B}$, ${\cal C}$ are defined by equations
$(B1)$, $(B2)$, $(B3)$ in Appendix B. We emphasize that $|m|=1$ is
no longer a solution with unconstrained $D^2$ for a partial MSID.
One may solve equation (3.2.11) for two values of $a^2D^2$
\begin{eqnarray}
a^2D^2=({\cal B}-{\cal A}-{\cal C}\pm\Delta^{1/2})
/[2(2-m^2-F|m|)]\ ,\quad\!  (3.2.12) \nonumber
\end{eqnarray}
where the determinant $\Delta$ is positive for $|m|\ge 1$ (see
definition [$B4$] in Appendix B).

For the case of $|m|=1$, criterion (3.2.7) becomes
$$
(a^2D^2-C_A^2/2)(1-F)(a^2D^2+a^2-C_A^2/2)=0\ ,\eqno(3.2.13)
$$
where $0<F<1$. As $2\pi Gr\Sigma_{0}=F[a^2(1+D^2)-C_A^2/2]>0$ is a
necessary requirement, the only possible solution of $a^2D^2$ for
a partial MSID becomes $ a^2D^2=C_A^2/2\ .
$

\subsection{Onset of bar-type instabilities in MSIDs}

For the disk stability problem, there exists a key parameter for
both secular and dynamic bar-type instabilities, namely, the ratio
of the kinetic energy of rotation ${\cal T}$ to the absolute value
of the gravitational potential energy ${\cal W}$ (Ostriker \&
Peebles 1973; Ostriker 1978). If one regards the stationary
configuration of aligned $|m|=2$ MSID as the onset of secular
bar-type instabilities (Shu et al. 2000), it is then of interest
to properly estimate the ratio ${\cal T}/|{\cal W}|$. For an MSID
of infinite radial extent, both quantities ${\cal T}$ and ${\cal
W}$ are infinite but their ratio remains finite. Starting from the
radial force balance of the background magnetorotational
equilibrium
$$
-\Sigma_{0}\Omega^2r=-d\Pi/dr -\Sigma_{0}\partial\phi_T/\partial r
-\Sigma_{0}C_A^2/(2r)\ ,\eqno(3.3.1)
$$
we multiply equation (3.3.1) by $2\pi r^2dr$ and integrate from
$0$ to a radius $R$ which is allowed to approach infinity
eventually. After an integration by parts for the gas pressure
term up to $R$, one obtains the MSID virial theorem
$$
2({\cal T}+{\cal U})+{\cal W}-{\cal M}=2\pi R^2\Pi (R)\
\eqno(3.3.2)
$$
within $R$, where by using equation (2.2.2) for $\Sigma_{0}$
$$
\! {\cal T}\equiv {1\over 2}\!\!\int_{0}^R\!\!\Sigma_{0}
r^2\Omega^22\pi rdr ={a^4D^2F\over 2G} \bigg[1+D^2-{C_A^2\over
2a^2}\bigg]R \eqno(3.3.3)
$$
is the kinetic energy of the MSID rotation,
$$
{\cal U}\equiv \int_{0}^R\Pi 2\pi rdr ={a^2F\over G}
\bigg[a^2(1+D^2)-{C_A^2\over 2}\bigg]R\ \eqno(3.3.4)
$$
is the gas internal energy,
$$
\! {\cal W}\equiv -\int_{0}^R\!\!\! r{d\phi_T\over
dr}\Sigma_{0}2\pi rdr =-{a^4F\over G} \bigg[1+D^2-{C_A^2\over
2a^2}\bigg]^2\!\! R
 \eqno(3.3.5)
$$
is the total gravitational potential energy (if there is no matter
outside $R$),
$$
\! {\cal M}\equiv
\int_{0}^R{\int dzB_{\phi}^2\over 8\pi}2\pi rdr ={C_A^2a^2F\over
2G}\bigg[1+D^2-{C_A^2\over 2a^2}\bigg]\! R \eqno(3.3.6)
$$
is the MSID magnetic energy within $R$.
The usual key ratio is then
$$
{\cal T}/|{\cal W}|=a^2D^2/[2 (a^2+a^2D^2-C_A^2/2)]\ .
\eqno(3.3.7)
$$
In view of the MHD virial theorem (3.3.2) in which ${\cal W}<0$ by
equation (3.3.5), one might suspect the quantity ${\cal W}-{\cal
M}$ to play the role of ${\cal W}$ in the absence of coplanar
magnetic field. Because
$$
{\cal W}-{\cal M}=-{a^2(1+D^2)F\over G}
\bigg[a^2(1+D^2)-{C_A^2\over 2}\bigg]R <0\ \eqno(3.3.8)
$$
for $\Sigma_{0}>0$, it would be suggestive of a new modified ratio
$$
{\cal T}/|{\cal W}-{\cal M}|=D^2/[2(1+D^2)]\ \eqno(3.3.9)
$$
to determine the instability criterion of an MSID for
nonaxisymmetric aligned MHD perturbations.

\subsection{The unaligned case of stationary
nonaxisymmetric MSID configurations}

We now consider the unaligned case for logarithmic spiral
structures (Kalnajs 1973; Shu et al. 2000). Combinations of
azimuthal momentum equation (2.4.3) with radial momentum equation
(2.4.2) and with mass conservation (2.4.1) give
\begin{eqnarray}
& & \bigg\lbrace m^2\Omega^2r^2-\kappa^2r^2+C_A^2 \nonumber \\ & &
-C_A^2\bigg[m^2-\bigg({C_A^2\over 2\Omega^2r^2} -{3\over
2}\bigg)\bigg({C_A^2\over 2\Omega^2r^2}-2\bigg) \bigg]\bigg\rbrace
iU \nonumber \\ & & =m\Omega r^2{\partial\Phi\over\partial r}
+2m\Omega r\Phi-{m\Omega C_A^2rS\over 2\Sigma_{0}} \nonumber \\ &
& \quad\qquad +C_A^2\bigg[r{\partial\over\partial r}
\bigg({m\Omega rS\over\Sigma_{0}} -{m\Phi\over\Omega r}\bigg)
\nonumber \\ & & \quad\qquad -\bigg({C_A^2\over
2\Omega^2r^2}-{5\over 2}\bigg) \bigg({m\Omega rS\over\Sigma_{0}}
-{m\Phi\over\Omega r}\bigg)\bigg]\ \qquad (3.4.1) \nonumber
\end{eqnarray}
and
$$
{m\Omega rS\over\Sigma_{0}} +r{\partial (iU)\over\partial r}
+\bigg({C_A^2\over 2\Omega^2r^2} -{\kappa^2\over
2\Omega^2}\bigg)iU -{m\Phi\over\Omega r}=0\ .\eqno(3.4.2)
$$
The Poisson integral (2.3.11) is a linear integral equation and
has a number of complete sets of eigenfunctions, depending on the
radial range of nonzero surface mass density (Snow 1952). For a
radial domain of $(0,1)$, the radial part of the eigenfunctions
may be Legendre polynomials (Hunter 1963) or Bessel functions
(Yabushita 1966), while for a radial domain of $(0,\infty)$, one
may use Bessel functions (Toomre 1963) or logarithmic spirals
(Kalnajs 1965, 1971). We here study unaligned logarithmic spiral
structures.

For logarithmic spirals (Kalnajs 1965, 1971) with constant pitch
angles, the known potential-density pair is
$$
S=sr^{-3/2+i\alpha}\ \ \hbox{ and } \ \ V=vr^{-1/2+i\alpha}\
,\eqno(3.4.3),\ (3.4.4)
$$
where $\alpha$ is a constant parameter that characterizes the
radial variation of perturbations and the two constant
coefficients $s$ and $v$ are related by
$$
v=-2\pi G{\cal N}_m(\alpha)s\ ,\eqno(3.4.5)
$$
\begin{eqnarray}
& &\!\!\!\!\!\!\!\!\!\!\!\!\hbox{with  }\ {\cal N}_m(\alpha)\equiv
K(\alpha,m) \nonumber \\ & & ={1\over
2}{\Gamma[(m+1/2+i\alpha)/2]\Gamma[(m+1/2-i\alpha)/2]
\over\Gamma[(m+3/2+i\alpha)/2]\Gamma[(m+3/2-i\alpha)/2] } \quad\ \
(3.4.6)\nonumber
\end{eqnarray}
being the Kalnajs function (Section IV of Kalnajs 1971) and
$\Gamma(...)$ being the gamma-function (Qian 1992; Appendix A of
Syer \& Tremaine 1996). The Kalnajs function is real and positive.
Table 1 of Kalnajs (1971) contains numerical values of
$K(\alpha,m)$ for different values of $\alpha$ and $m$. For large
$m$ or large $\alpha$, one has ${\cal N}_m(\alpha)\cong
(m^2+\alpha^2)^{-1/2}$ approximately. For $m=0,1,2$, the curves of
$K(\alpha,m)$ versus $\alpha$ were displayed in Figure 1 of Shu et
al. (2000). The Kalnajs function is even in $m$
and $\alpha$.
With a sufficient accuracy (Shu et al. 2000), one may
approximately take
$$
{\cal N}_m(\alpha)\cong(m^2+\alpha^2+1/4)^{-1/2}\ .\eqno(3.4.7)
$$
The exact recurrence or recursion relation for ${\cal
N}_m(\alpha)$ is
$$
{\cal N}_{m+1}(\alpha){\cal N}_m(\alpha)
=[(m+1/2)^2+\alpha^2]^{-1}\ \eqno(3.4.8)
$$
(Kalnajs 1971). Suppose one takes, for example,
$$
{\cal N}_2(\alpha)\cong (4+\alpha^2+1/4)^{-1/2}\
$$
by approximation (3.4.7) with $|m|=2$, it then follows
successively from the recursion (3.4.8) that
$$
{\cal N}_1(\alpha)\cong
=(1+1/16+\alpha^2/4)^{1/2} /(1+1/8+\alpha^2/2)\ <\ 1
$$
and
$$
{\cal N}_0(\alpha)\cong {(1+1/8+\alpha^2/2) \over
(\alpha^2+1/4)(1+1/16+\alpha^2/4)^{1/2}}\
$$
(see
Fig. 1 of Shu et al. 2000). To achieve a higher accuracy for
${\cal N}_m(\alpha)$, one starts from approximation (3.4.7) with a
large $|m|$ and use the recursion (3.4.8) to successively derive
${\cal N}_m(\alpha)$ of descending $|m|$.

Consistent with the logarithmic spiral forms of $S(r)$ and $V(r)$
by equations (3.4.3) and (3.4.4), we take $U$ to be
$$
iU=iur^{-1/2+i\alpha}\ ,\eqno(3.4.9)
$$
where $u$ is a constant coefficient. It follows from equations
(2.4.1), (2.4.2), (3.4.5), (2.3.12), and (2.3.15) that
$$
{m\Omega s\over\Sigma_{0}} +iu\bigg(i\alpha+{C_A^2\over
2\Omega^2r^2}-{3\over 2}\bigg) -{m\over\Omega r}\bigg({a^2s\over
r\Sigma_{0}}+v\bigg)=0\ , \eqno(3.4.10)
$$
\begin{eqnarray}
& & \!\!\!\!\!\!\!\!\!\!\!\! \bigg\lbrace
m^2\Omega^2r^2-\kappa^2r^2-C_A^2\bigg[m^2-1 \nonumber \\ &
&\!\!\!\!
-\bigg({C_A^2\over 2\Omega^2r^2}-{3\over 2}\bigg)
\bigg({C_A^2\over 2\Omega^2r^2}-2\bigg)\bigg]\bigg\rbrace iu
\nonumber \\ & &\!\!\!\!\!\!\!\!\!\!\!\! =C_A^2\bigg(i\alpha +2
-{C_A^2\over 2\Omega^2r^2}\bigg) \bigg[{m\Omega
s\over\Sigma_{0}}-{m\over\Omega r} \bigg({a^2s\over
r\Sigma_{0}}+v\bigg)\bigg] \nonumber \\ & & +m\Omega
r(i\alpha+3/2)\bigg({a^2s\over r\Sigma_{0}}+v\bigg) -{m\Omega
C_A^2s\over 2\Sigma_{0}} \ ,\ \qquad (3.4.11)\nonumber
\end{eqnarray}
$$
iR={r^{1/2}B_{\theta}\over\Omega r}iu r^{-1+i\alpha}
={B_{\theta}iU\over\Omega r}\ ,\eqno(3.4.12)
$$
$$
Z=-{i\alpha r^{1/2}B_{\theta}\over m\Omega r}iu r^{-1+i\alpha}
=-{i\alpha B_{\theta}\over m\Omega r}iU\ .\eqno(3.4.13)
$$
Equation (3.4.10) relates $s$ and $iu$ by relation (3.4.5).
Independently, equation (3.4.11) is another relation for $s$ and
$iu$ by relation (3.4.5). A combination of the two resulting
relations in terms of $s$ and $iu$ then gives rise to the solution
criterion for stationary logarithmic spirals in an MSID. With an
exact cancellation of the imaginary part, the {\it solution
criterion} or {\it dispersion relation} for stationary unaligned
logarithmic spirals becomes
\begin{eqnarray}
& & \!\!\!\!\!\!\!\!\!\!\!\!\! \bigg({a^2s\over
r\Sigma_{0}}+v\bigg) \bigg[{3\over 2}\bigg({C_A^2\over
2\Omega^2r^2}-{3\over 2}\bigg) -\alpha^2\bigg]-{C_A^2s\over
2\Sigma_{0}r} \bigg({C_A^2\over 2\Omega^2r^2}-{3\over 2}\bigg)
\nonumber \\ & & \quad =\bigg[ m\Omega r-{2\Omega r\over m}
-C_A^2\bigg({\alpha^2-1\over m\Omega r} +{m\over\Omega
r}\bigg)\bigg] \nonumber \\ & &\qquad\qquad
\times\bigg[{m\over\Omega r}\bigg({a^2s\over r\Sigma_{0}}+v\bigg)
-{m\Omega s\over\Sigma_{0}}\bigg]\qquad\qquad\ (3.4.14) \nonumber
\end{eqnarray}
where $s$ and $v$ are related by equation (3.4.5). A physically
more revealing form of criterion (3.4.14) is
\begin{eqnarray}
& & \!\!\!\!\!\!\!\!\!\!\! m^4\Omega^4-\lbrace 2\Omega^2
+[C_A^2/r^2+a^2/r^2-2\pi G{\cal N}_m(\alpha)\Sigma_{0}/r ]
\nonumber \\ & &\!\!\!\!\!\!\!\!\!\!\!\quad \times
(m^2+\alpha^2+1/4)-2C_A^2/r^2\rbrace m^2\Omega^2 \nonumber \\ &
&\!\!\!\!\!\!\!\!\!\!\!\qquad +(m^2C_A^2/r^2)\lbrace [a^2/r^2-2\pi
G{\cal N}_m(\alpha)\Sigma_{0}/r ] \nonumber \\ &
&\!\!\!\!\!\!\!\!\!\!\!\qquad \times
(m^2+\alpha^2+1/4-1/2)-C_A^2/(4r^2)\rbrace=0\ .\ \quad(3.4.15)
\nonumber
\end{eqnarray}
In parallel, the two stationary solution criteria (3.2.6) and
(3.4.15) for the {\it aligned} and {\it unaligned} cases
correspond to each other remarkably well, especially in view of
the effective wavenumber $(m^2+\alpha^2+1/4)^{1/2}r^{-1}$ (Shu et
al. 2000). Again, the form of equation (3.4.15) reminds us of the
dispersion relation for spiral FMDWs and SMDWs (Lou 1996; Fan \&
Lou 1996; Lou \& Fan 1998a; LYF 2001). These logarithmic spiral
MHD density waves propagate in both radial and azimuthal
directions relative to the MSID. For stationary logarithmic
spirals in an inertial frame of reference, criterion (3.4.15)
leads to two possible values of $a^2D^2$. As both $\kappa^2\equiv
2\Omega^2$ and $\Sigma_{0}$ contain the $D^2$ parameter, the
determination of $D^2$ should be somewhat different from the
standard WKBJ wave results derived in a background that is
prescribed {\it a priori}. For example, for a full SID that is
isopedically magnetized, Shu et al. (2000) obtains only one real
solution of $\alpha$ at a given $D^2$ for each $m\ge 1$ as shown
in their Fig. 3 instead of the two branches of long- and
short-waves (see also Syer \& Tremaine 1996 for a discussion).
While for a sufficiently large $D^2$ value in the special case of
$m=0$, there are two values of $\alpha$ as shown in their Fig. 2,
bordering the ring fragmentation regime.

Equation (3.4.15) may be cast into the form of
\begin{eqnarray}
& & [a^2-\Omega^2r^2-2\pi G{\cal N}_m(\alpha)\Sigma_{0} r]
\nonumber \\ & &\ \ \times [(m^2+\alpha^2
+1/4)\Omega^2r^2-(m^2+\alpha^2-1/4)C_A^2] \nonumber \\ &
&\quad\qquad =-\alpha^2\Omega^4r^4-(C_A^2/2-3\Omega^2r^2/2)^2\ .\
\qquad (3.4.16)\nonumber
\end{eqnarray}
After a straightforward rearrangement using the $\Sigma_{0}$
profile (2.2.2), criterion (3.4.16) becomes
\begin{eqnarray}
& & \!\!\! [ m^2 +\alpha^2+1/4 -(m^2+\alpha^2-1/4)C_A^2/(a^2D^2)]
\nonumber \\ & &\!\!\!\quad \times\lbrace
a^2-[(1+D^2)a^2-C_A^2/2]F{\cal N}_m(\alpha)\rbrace \nonumber \\ &
&\!\!\! \quad\quad -a^2D^2[m^2-2-(m^2+\alpha^2-1)C_A^2/(a^2D^2)]
\nonumber \\ & &\!\!\!\qquad\quad\quad
+(C_A^2/2)[C_A^2/(2a^2D^2)-3/2]=0\ . \qquad (3.4.17) \nonumber
\end{eqnarray}
For $C_A^2=0$ and $F=1$, this criterion reduces to equation (37)
of Shu et al. (2000).
Also with $C_A^2=0$ and $\alpha^2\rightarrow 0$ in the so-called
``breathing mode" regime, one obtains
$$
m^2D^2=2D^2+(m^2+1/4)[1-(1+D^2)F{\cal N}_m(0)]\ ,\eqno(3.4.18)
$$
which differs from condition (3.2.2) derived earlier for the
aligned case, especially in the regime of large $|m|$.

In terms of a quadratic algebraic equation for $a^2D^2$, equation
(3.4.17) may be written in the explicit form of
\begin{eqnarray}
& & \!\!\!\!\!\!\!\!\!\!\!\!\! [m^2-2+(m^2+\alpha^2+1/4)F{\cal
N}_m(\alpha)]a^4D^4 \qquad\qquad\ \ (3.4.19)\nonumber \\ & &
\!\!\!\!\!\!\!\!\!\!\! -({\cal A}+{\cal B}+{\cal C})a^2D^2
+{C_A^2(m^2+\alpha^2-1/4){\cal A}\over (m^2+\alpha^2+1/4)}
-{C_A^4\over 4}=0\ ,\nonumber
\end{eqnarray}
where ${\cal A}$, ${\cal B}$, ${\cal C}$ are defined by equations
$(C1)$, $(C2)$, $(C3)$ in Appendix C. For $|m|\ge 2$ in equation
(3.4.19), the coefficient of $a^4D^4$ is positive, the coefficient
of $a^2D^2$ is negative, and the remaining coefficient is
positive.
By the definition of ${\cal N}_1(\alpha)$, this statement remains
valid with $|m|=1$ for the first two coefficients and for the last
coefficient with an {\it additional sufficient} requirement
$a^2>C_A^2/4$.
The determinant $\Delta$ of the quadratic equation (3.4.19) is
positive for $|m|\ge 2$ such that there are two positive $a^2D^2$
(see Appendix C).

One can further show, in a similar manner, that the determinant
$\Delta$ of equation (3.4.19) is non-negative for $F=1$ and
$|m|=1$ with {\it sufficient} requirements $\alpha\ge\sqrt{3}/2$
and $a\ge C_A/2$. In other words, there are two positive values of
$D^2$ as long as $\alpha\ge\sqrt{3}/2$ and $a\ge C_A/2$.

For $F=1$, $m=0$, and a $\alpha$ greater than a specific value
$\alpha_c$ in equation (3.4.19), one can show that the coefficient
of $a^4D^4$ is positive, the coefficient of $a^2D^2$ is negative,
and the remaining coefficient is positive. Following the same
procedure of proof, one can further show that there exist two
positive roots of $a^2D^2$. For $0\le \alpha\le\alpha_c$, however,
the solution structure of condition (3.4.19) may have several
possiblities, depending on the ratio $q^2\equiv C_A^2/a^2$. In
terms of $q^2$, $\alpha$, and ${\cal N}_{0}(\alpha)$, the two
solutions of $D^2$ are
\begin{eqnarray}
& & D^2= \Bigg[\!\!\!\Bigg[ \{[1-(1-q^2/2){\cal
N}_0(\alpha)](\alpha^2+1/4) \nonumber \\ & &\qquad
+q^2(\alpha^2-1/4){\cal N}_0(\alpha)+q^2(\alpha^2-7/4) \}
\nonumber \\ & & \pm \bigg[\{[1-(1-q^2/2){\cal
N}_0(\alpha)](\alpha^2+1/4) \nonumber \\ & &\qquad
+q^2(\alpha^2-1/4){\cal N}_0(\alpha)+q^2(\alpha^2-7/4) \}^2
\nonumber \\ & &\qquad -4[-2+(\alpha^2+1/4){\cal
N}_0(\alpha)]\lbrace q^2(\alpha^2-1/4) \nonumber \\ & &\qquad
\times [1-(1-q^2/2){\cal N}_0(\alpha)]
-q^4/4\rbrace\bigg]^{1/2}\Bigg]\!\!\!\Bigg] \nonumber \\ & &
\qquad\qquad \times \lbrace 2[-2+(\alpha^2+1/4){\cal
N}_0(\alpha)]\rbrace^{-1}\ , \qquad (3.4.20)\nonumber
\end{eqnarray}
where the denominator changes from negative to positive between
$\alpha\sim 1.79$ and 1.795 approximately\footnote{We used an
approximate expression of ${\cal N}_0(\alpha)$. Shu et al. (2000)
derived a value of $|\alpha_c|=1.759$ from the same condition
$(\alpha_c^2+1/4){\cal N}_0(\alpha_c)=2$.}; this value of
$\alpha$, independent of $q^2$, is denoted by $\alpha_c$. We shall
refer to the two values of $D^2$ as the plus- and minus-sign
solutions according to their respective signs in front of the
square root of the determinant in equation (3.4.20). As specific
examples, we explored numerically three cases with decreasing
values of ratio $q^2$, namely, (a) $q^2=3.61$ (Fig. 1a), (b)
$q^2=1.0$ (Fig. 1b), and (c) $q^2=0.09$ (Fig. 1c). For
$\alpha>\alpha_c$, there are indeed two positive values of $D^2$
as stated earlier. The larger and smaller ones correspond to the
plus- and minus-sign solutions, respectively. As
$\alpha\rightarrow\alpha_c+0^{+}$, the plus-sign solution goes to
$+\infty$, while the minus-sign solution remains finite and
continuous across $\alpha_c$. For $q^2\neq 0$, there exists a
finite interval $\alpha_l<\alpha<\alpha_u$ such that the
determinant $\Delta$ is {\it negative} and there are thus no real
solutions for $D^2$. Approximately, $\alpha_l$ is between 0.595
and 0.6 while $\alpha_u$ is between 0.815 and 0.82 for $q^2=3.61$;
$\alpha_l$ is between 0.68 and 0.685 while $\alpha_u$ is between
0.92 and 0.925 for $q^2=1.0$; and $\alpha_l$ is between 0.96 and
0.965 while $\alpha_u$ is between 1.08 and 1.085 for $q^2=0.09$.
There is a systematic shift of this open interval
$(\alpha_l,\alpha_u)$ towards larger $\alpha$ as $q^2$ decreases.
For $\alpha_u\le\alpha\le\alpha_c$, the plus-sign solution is
negative while the minus-sign solution changes from negative to
positive for increasing $\alpha$. For $0\le\alpha\le\alpha_l$, the
minus-sign solution is greater than the plus-sign solution and the
two solutions join at $\alpha=\alpha_l$ with a $D^2>0$; for
$q^2=1$ and $q^2=0.09$, the smaller plus-sign solutions change
from negative to positive as $\alpha$ increases from $\alpha=0$
within this interval. All negative values of $D^2$ are not shown
in Fig. 1 as they are unphysical.

There are now two ring fragmentation regimes when $\alpha$ is
sufficiently large (one for larger $D^2$ and one for smaller
$D^2$) and one modified collapse regime for sufficiently small
$\alpha$ and $D^2$ as shown in Figure 1 for all three $q^2$
values. In the limit of $q^2\rightarrow 0$, the situation
degenerates to the case shown in Figure 2 of Shu et al. (2000) as
expected (see also Lemos et al. 1991 on the linear stability of
axisymmetric scale-free disks to axisymmetric disturbances). For
increasing values of $q^2$, the lower ring fragmentation regime
associated with slow MHD disturbances is enlarged while the upper
ring fragmentation regime associated with fast MHD disturbances is
pushed upwards in the parameter scheme of Fig. 1. That is, one
needs an even larger $D^2$ to access the upper ring fragmentation
regime.

As the MHD generalization of Toomre's $Q$-parameter, the
$Q_M$-parameter (Lou \& Fan 1998a) is defined by
$$
Q_M\equiv {(a^2+C_A^2)^{1/2}\kappa\over\pi G\Sigma_0}
={2\sqrt{2}D(1+q^2)^{1/2}\over F(1+D^2-q^2/2)}\eqno(3.4.21)
$$
with an MSID profile (2.2.2). For axisymmetric FMDWs, the disk
stability requires $Q_M>1$ (Lou \& Fan 1998a). For the minima
$D_{min}^2$ of the boundaries of the upper fragmentation regime
and the relevant values of $q^2$, the corresponding values of
$Q_M$ thus obtained are fairly close to unity in all three cases
of Figure 1. For example, with $q^2=3.61$, $D^2_{min}=40.62$
at $\alpha=3.9$ and $Q_M=0.972$; with $q^2=1.0$, $D^2_{min}=15.24$
at $\alpha=3.681$ and $Q_M=0.992$; and with $q^2=0.09$,
$D^2_{min}=6.48$
at $\alpha=3.264$ and $Q_M=1.01$. It appears that the critical
$Q_M$ value tends to increase slightly with decreasing values of
$q^2$ and with decreasing values of critical $\alpha$. This trend
is also complemented by the limiting case (Shu et al. 2000) with
$q^2=0$, $D_{min}^2=5.41$ at $\alpha=3.056$, and $Q_M=1.026$. Thus
the approximate instability criterion $Q_M\lsim 1$ (Fan \& Lou
1996; Lou \& Fan 1998a) appears pertinent to the upper branch of
ring fragmentation for axisymmetric fast MHD disturbances (see
Fig. 1).

Comparing with Fig. 2 of Shu et al. (2000), the coplanar magnetic
field modifies the collapse regime as shown in our Fig. 1 and
introduces the lower regime of ring fragmentation for slow MHD
disturbances when $\alpha$ is sufficiently large. These
instabilities do not occur when $D^2$ is large: $D^2\gsim 9.2$ for
$q^2=3.61$; $D^2\gsim 3.5$ for $q^2=1$; $D^2\gsim 1.2$ for
$q^2=0.09$; and $D^2\gsim 0.932$ for $q^2=0$. The last result of
$q^2=0$ is taken from subsection 4.1 of Shu et al. (2000).

From equation (3.4.13)
and
$$
iU=\bigg({m\Phi\over\Omega r}-{m\Omega rS\over\Sigma_{0}}\bigg)
\bigg(i\alpha+{C_A^2\over 2\Omega^2r^2}-{3\over 2}\bigg)^{-1}\ ,
\eqno(3.4.22)
$$
we derive
\begin{eqnarray}
& & Z=-{i\alpha B_{\theta}S\over\Omega^2 r} \bigg[{a^2\over
r\Sigma_{0}}-2\pi G{\cal N}_m(\alpha)
-{\Omega^2r\over\Sigma_{0}}\bigg] \nonumber \\ & &
\qquad\qquad\qquad \times\bigg(i\alpha+{C_A^2\over
2\Omega^2r^2}-{3\over 2}\bigg)^{-1}
\ ,\qquad\quad\ (3.4.23) \nonumber
\end{eqnarray}
which is important to determine the spatial phase relationship
between the surface mass density and azimuthal magnetic field
perturbations. In the limit of $\alpha\rightarrow 0$, one has
$Z\rightarrow 0$, but $iU$ given by (3.4.22) remains nonzero. For
a small $\alpha\neq 0$ such that $\alpha^2$ may be dropped
relative to $i\alpha$ term, $Z$ and $S$ are phase shifted by
$\sim\pm\pi/2$.

In the limit of $\alpha\rightarrow\infty$, we obtain
$$
Z=-{B_{\theta}S\over\Sigma_{0}\Omega^2r^2} [a^2-2\pi G{\cal
N}_m(\alpha)\Sigma_{0}r-\Omega^2r^2]\ \eqno(3.4.24)
$$
which, given stationary condition (3.4.16), can be cast into a
convenient form to examine the spatial phase relationship between
$Z$ and $S$ in an MSID plane at $z=0$.

\section{Phase Relationships between Magnetic FieLd and Mass Density }

We examine spatial phase relationships among velocity
disturbances, azimuthal magnetic field, and surface mass density
enhancements because they provide useful observational diagnostics
for magnetized spiral galaxies (Beck \& Hoernes 1996; Fan \& Lou
1996; Lou \& Fan 1998a, 2002; Frick et al. 2000). Regions of
high-density gas are vulnerable to active star formation, while
nonthermal radio-continuum emissions from gyrating relativistic
cosmic-ray electrons trapped in a spiral galaxy would reveal
regions of stronger magnetic field. Thus, large-scale spiral
structures of optical and radio-continuum emissions contain
valuable information of the underlying MHD (Lou \& Fan 2000a, b).
The mathematical development is somewhat lengthy and the relevant
formulae are summarized in Appendices D and E for aligned and
unaligned perturbations respectively. A reader may want to mainly
concentrate on the flow of logics with convenient references to
equations in the two Appendices.

\subsection{Aligned MSID Configurations}

For a full MSID of $F=1$, $D^2$ as given by solution (3.2.10)
depends only on two dimensionless parameters $q^2\equiv C_A^2/a^2$
and $|m|$. By solutions (3.2.10), one derives equation $(D1)$ that
shows $a^2D^2-C_A^2/2>0$ for the upper {\it plus-sign} solution in
equation $(D1)$. For the lower {\it minus-sign} solution in
equation $(D1)$, whether inequality $a^2>C_A^2|m|/[2(|m|-1)]$
holds or not in the determinant $\Delta$ would determine whether
$a^2D^2-C_A^2/2>0$ or not.

By equation $(D1)$, we then examine the sign of $2\pi
G\Sigma_{0}r=F(a^2+a^2D^2-C_A^2/2)$ for a full MSID of $F=1$. An
addition of $a^2$ to the first term on the RHS of equation $(D1)$
gives equation $(D2)$. Examine then the determinant $\Delta$ in
solutions (3.2.10) or $(D1)$ in the form of equation $(D3)$. The
sum of the last two terms in expression $(D3)$ is given by
expression $(D4)$. When $a^2>C_A^2/4$, we have the background
surface mass density $\Sigma_{0}>0$ for both the plus- and
minus-sign solutions (3.2.10) of $a^2D^2$. For $a^2<C_A^2/4$, one
has $\Sigma_0>0$ for the plus-sign solution only; the minus-sign
solution gives $\Sigma_0<0$ which is unphysical.

To examine the spatial phase relationship between $b_{\theta}$ and
$\Sigma_1$ in the MSID plane, we use relations (3.1.8) and (3.1.9)
derived in subsection 3.1. Specifically, we determine the signs of
both numerator and denominator on the RHS of equation (3.1.9). For
the numerator $2\Omega^2r^2-C_A^2/2=2(a^2D^2-C_A^2/2+C_A^2/4)$, we
add $C_A^2/4$ to the first term on the RHS of equation $(D1)$ to
derive equation $(D5)$. The determinant $\Delta$ may be cast into
the form of equation $(D6)$. For $a^2-C_A^2/4>0$ in equation
$(D6)$, one has both $\Sigma_{0}>0$ and $2a^2D^2-C_A^2/2>0$ for
either plus- or minus-solutions in equation (3.2.10). For
$a^2-C_A^2/4<0$, only the plus-sign solution with
$a^2D^2-C_A^2/4>0$ is physically valid.

For the sign of the denominator on the RHS of equation (3.1.9) as
given by equation $(D7)$, a subtraction of $C_A^2-C_A^2/(2m^2)$
from the first term on the RHS of solution (3.2.10) leads to
expression $(D8)$. We now examine the determinant $\Delta$ in the
form of expression $(D9)$. The last term containing $(3|m|^2-2)$
on the RHS of equation $(D9)$ is positive for $|m|\neq 0$. Hence,
the denominator $m^2\Omega^2r^2-C_A^2(m^2-1/2)$ is positive and
negative for the plus- and minus-sign solutions in equation
(3.2.10), respectively.

To examine the sign of the background surface mass density
$\Sigma_{0}$ of a partial MSID for $|m|\ge 1$, we consider
$a^2D^2+a^2-C_A^2/2$. By adding $a^2-C_A^2/2$ to the term
involving ${\cal B}-{\cal A}-{\cal C}$ on the RHS of solution
(3.2.12) for $a^2D^2$, one has inequality $(D10)$. In the
determinant $\Delta$ of solution (3.2.12), we then consider the
part as given by equation $(D11)$. When $a^2-C_A^2/4>0$,
$\Sigma_0$ is positive for solutions (3.2.12) of both signs; for
$a^2-C_A^2/4<0$, $\Sigma_0$ is positive for the plus-sign
solution, while $\Sigma_0<0$ for the minus-sign solution. These
conclusions for a partial MSID ($0\le F<1$) turn out exactly the
same as those for a full MSID ($F=1$).

To examine the phase relationship between $Z$ and $S$, equations
(3.1.7)$-$(3.1.9) remain valid with $\Sigma_0$ profile (2.2.2) for
a partial MSID. Again, we examine the signs of the numerator and
denominator on the RHS of (3.1.9). For the numerator, we consider
$a^2D^2-C_A^2/4$ by subtracting $C_A^2/4$ from the term involving
${\cal B}-{\cal A}-{\cal C}$ on the RHS of solution (3.2.12) for
$a^2D^2$. The resulting expression is then given by equation
$(D12)$. In the determinant $\Delta$ of solution (3.2.12), we then
consider the part given by equation $(D13)$. Therefore,
$\Sigma_{0}$ and $a^2D^2-C_A^2/4$ are positive for both plus- and
minus-sign solutions of (3.2.12) when $a^2-C_A^2/4>0$. For
$a^2-C_A^2/4<0$, only the plus-sign solution is valid with
$a^2D^2-C_A^2/4>0$; the minus-sign gives a $D^2<0$.

We now examine the sign of the denominator on the RHS of equation
(3.1.9). For this purpose, we consider $a^2D^2-[1-1/(2m^2)]C_A^2$.
In the determinant $\Delta$ of solution (3.2.12), we consider the
portion given by equation $(D14)$, which vanishes for $|m|=1$ and
is positive for $|m|\ge 2$. Thus, $m^2\Omega^2r^2-C_A^2(m^2-1/2)$
is positive and negative for the plus- and minus-sign solutions of
(3.2.12), respectively. Using these results in equations (3.1.8)
and (3.1.9), we have $Z$ and $S$ being out of phase for the
plus-sign solution of (3.2.12), and $Z$ and $S$ being in phase for
the minus-sign solution of (3.2.12) when $a^2>C_A^2/4$. When
$a^2<C_A^2/4$, the minus-sign solution of (3.2.12) leads to a
negative $\Sigma_{0}$.

Based on these analyses and in reference to equations (3.1.8) and
(3.1.9) for the phase relationships between $Z$ and $S$,
perturbation enhancements of $b_{\theta}$ and $\Sigma_1$
anticorrelate and correlate with each other for the plus- and
minus-sign solutions, respectively, when $a^2>C_A^2/4$. When
$a^2<C_A^2/4$, the minus-sign solution is invalid because the
surface mass density $\Sigma_0$ would be negative. These two
distinct stationary aligned MSID configurations appear as results
of differential rotation, self-gravity, and curved magnetic field.
According to equation (3.1.4), one may write
$$
iU={m\Omega rS\over\Sigma_{0}} {(a^2-2\pi G\Sigma_{0}
r/|m|-\Omega^2r^2) \over (C_A^2/2-\Omega^2r^2)}\ .\eqno(4.1.1)
$$
Therefore, $iU\propto \pm rS$, $iR\propto \pm rS$, and $J\propto
-S$ for the plus- and minus-sign solutions. With these spatial
phase relationships among stationary perturbation variables, one
may conceive mental pictures for the two distinct types of
stationary MSID bar configurations for $|m|=2$. For the plus-sign
case, enhancements of $b_{\theta}$ and $\Sigma_1$ are out of phase
with each other. High-density gas regions are active in star
formation (i.e., bright in optical and infrared bands) with
enhanced small-scale random magnetic fields (i.e., bright in total
nonthermal radio-continuum emissions), whereas relatively strong
regular magnetic field regions should be bright in polarized
radio-continuum emissions with higher degrees of polarization
(less disturbed by activities associated with cloud and star
formation on small scales). For the minus-sign case, enhancements
of $b_{\theta}$ and $\Sigma_1$ are in phase. By the same
rationale, the structural manifestations of bars or lopsided disks
in {\it total} and {\it polarized} radio emissions will be in
competition because small-scale random motions tend to enhance
total radio-continuum emissions while weaken polarized
radio-continuum emissions. Regions of strong total
radio-continuum, optical, infrared emissions should more or less
overlap. Depending on the level of activities in star-forming
regions, polarized radio-continuum emissions may also be
sufficiently strong in optically bright regions. Polarized
radio-continuum emissions from relatively weak magnetic field
regions should show higher degrees of polarization because of
reduced level star formation activities there.

\subsection{Unaligned Logarithmic MSID Spirals }

We first consider the important factor that appears in criterion
(3.4.16) for logarithmic spirals in an MSID, as given by equation
$(E1)$. By equations (3.4.16) and (3.4.23), the sign of this
factor $(E1)$ determines the spatial phase relationship between
$b_{\theta}$ and $\Sigma_1$. Using the stationary criterion in the
form of equation (3.4.19), we derive two values of $a^2D^2$ as
given by equation $(E2)$.
The determinant $\Delta$ under the square root of solution $(E2)$
can be shown to be non-negative for $|m|\geq 2$ (see Appendix C).
As $\Omega^2r^2=a^2D^2$, we subtract
$C_A^2\{1-[2(m^2+\alpha^2+1/4)]^{-1}\}$ from solution $(E2)$ and
rearrange the determinant $\Delta$ of solution $(E2)$ accordingly.
In the expression of determinant $\Delta$, we consider the
relevant part as given by equation $(E3)$, which is positive for
$|m|\ge 2$. For a full MSID of $F=1$ with $|m|=1$ and $|m|=0$, the
reader is referred to numerical results and discussions that
follow equation (3.4.19). It then follows with $|m|\ge 2$ that the
key factor $(E1)$ (see solution criterion [3.4.16]), namely $
(m^2+\alpha^2+1/4)\Omega^2r^2-(m^2+\alpha^2-1/4)C_A^2\ , $ is
positive and negative for {\it plus-} and {\it minus-sign}
solutions of $a^2D^2$ in equation $(E2)$, respectively.

Using dispersion relation (3.4.16) for stationary logarithmic
spirals in an MSID and relation (3.4.23) between $Z$ and $S$
derived in subsection 3.4, we finally arrive at
$$
Z\propto \pm S \{\alpha^2+i\alpha [C_A^2/(2\Omega^2r^2)-3/2]\}\
\eqno(4.2.1)
$$
for the plus- and minus-sign solutions of $a^2D^2$ in equation
$(E2)$, respectively. For a sufficiently large $\alpha$ in the
tight-winding or WKBJ regime, one may ignore the imaginary part in
relation (4.2.1). In this regime, $Z$ is approximately in-phase
and out-of-phase with $S$ for the plus- and minus-sign solutions,
respectively. This appears to be consistent with the results of
FMDW and SMDW analyses in the tight-winding approximation (Fan \&
Lou 1996; Lou \& Fan 1998a, 2002). For a sufficiently small
$\alpha$ that corresponds to relatively open spiral structures, we
may drop the $\alpha^2$ term in comparison with the term
proportional to $i\alpha$. In this regime of small $\alpha\neq 0$,
$Z$ is either ahead of or lag behind $S$ by a phase difference of
$\sim\pi/2$ for both stationary fast and slow logarithmic spiral
configurations in an MSID; this is a new result, not known before.

\section{Discussion and Applications }

\subsection{A discussion on bars and barred spirals}

The modal formulation and computations of Bertin et al. (1989a, b)
and Bertin \& Lin (1996) set a theoretical framework to classify
morphologies of spiral galaxies. Meanwhile, the analytical and
numerical results of Shu et al. (2000; also Galli et al. 2001)
provide an important perspective for the onsets of bar-type and
barred-spiral instabilities in isopedically magnetized SIDs
associated with nonaxisymmetric aligned disturbances and unaligned
logarithmic spiral perturbations that appear stationary in an
inertial frame of reference. One would like to know the overall
connection between the results and interpretations of Shu et al.
(2000) and those of Bertin et al. (1989a, b; Bertin \& Lin 1996),
especially regarding the cubic dispersion relation and accurately
solved numerical solutions in the modal scenario for bars and
barred spiral galaxies. Among several apparent differences in the
model formulations such as SIDs versus prescribed disks,
logarithmic spiral perturbations versus normal modes, stationarity
versus time variations, exact analytical solutions versus extended
WKBJ approximation as well as accurate numerical solutions, we
single out one key element that, we believe, distinguishes the two
independent lines of approach in a significant manner. The issue
involves analyses that are somewhat technical and subtle. We shall
describe them below.

For SIDs and their kin (Mestel 1963), there have been theoretical
studies on their structures and instability properties for decades
(Zang 1976; Toomre 1977; Lemos et al.
1991; Lynden-Bell \& Lemos 1993; Syer \& Tremaine 1996; Evans \&
Read 1998; Goodman \& Evans 1999; Shu et al. 2000; Galli et al.
2001) in view of their potential applications to the structure of
lopsided or normal and barred spiral galaxies, to the light cusps
seen in the nuclei of galaxies, and to the formation and collapse
of cloud cores in the birth of stars and planetary systems. Syer
\& Tremaine (1996) found semianalytic and numerical solutions of
nonaxisymmetric stationary equilibria for completely flattened
(razor-thin) power-law disks. The basic problem along with
pertinent issues have been well summarized by Shu et al. (2000),
and for the special case of index $\beta=0$ (in the notation of
Syre \& Tremaine 1996), Shu et al. (2000) derived analytic
solutions and criteria for both aligned and unaligned logarithmic
stationary perturbations in razor-thin isopedically magnetized
SIDs.

For aligned perturbations, the zero-frequency solutions correspond
to the onset of bifurcations from axisymmetric SIDs to
nonaxisymmetric SIDs in close analogy to bifurcations from
incompressible uniformly rotating Maclaurin spheroids to Dedekind
ellipsoids with configuration axes that remain fixed in space
(Chandrasekhar 1969; Binney \& Tremaine 1987). Moreover, Shu et
al. (2000) relate the aligned case of $m=2$ to the onset of the
secular barlike instability in the context of galactic dynamics
(Hohl 1971; Miller et al. 1970; Kalnajs 1972; Ostriker \& Peebles
1973; Bardeen 1975; Aoki et al. 1979; Vandervoort 1982, 1983).

In analyses of Feldman \& Lin (1973), Lau \& Bertin (1978), Lin \&
Lau (1979) and Bertin et al. (1989a, b) on galactic density waves,
there is a coefficient, usually referred to as $B$, in the
standard integro-differential equations for spiral density waves,
namely
$$
B=-{m^2\over r^2}-{4m\Omega (d\nu/dr)\over\kappa r(1-\nu^2)}
+{2m\Omega\over r^2\kappa\nu}{d\ln [\kappa^2/(\mu_{\circ}\Omega)]
\over d\ln r}\ \eqno(5.1)
$$
(see eq. [25b] of Lin \& Lau 1979 or eq. [2.5] of Bertin et al.
1989b), where $\nu^2\equiv (\omega-m\Omega)^2/\kappa^2$. The third
term on the RHS of equation $(5.1)$ is related to the corotation
resonance and the second term on the RHS of $(5.1)$ contains the
information of the $T_1$ or $J^2$ term proportional to
$d\ln\Omega/d\ln r$ with $J^2\equiv (2\pi
G\mu_{\circ}/\kappa^2)^2T_1$ (Lau \& Bertin 1978; Bertin et al.
1989b; cf. comments of Hunter 1983, Lou \& Fan 1998b, and MYE
1999). The second term on the RHS of $(5.1)$ may be split further
into the form of
$$
-{4m\Omega (d\nu/dr)\over\kappa r(1-\nu^2)}
={4m^2\Omega^2\over\kappa^2r^2(1-\nu^2)} {d\ln\Omega\over d\ln
r}+{4m\Omega\nu \over\kappa^2r(1-\nu^2)}{d\kappa\over dr}\
,\eqno(5.2)
$$
where the first item on the RHS is the $-T_1/(1-\nu^2)$ term (the
notation $T_1$ was introduced by Lau \& Bertin [1978]), or
equivalently, $-J^2/(1-\nu^2)$ term (the notation $J^2$ was later
used by Lin \& Lau [1979] and Bertin et al. [1989a, b]; see
footnote 2 of Lou \& Fan 1998b). To derive the cubic dispersion
relation of density waves, this $T_1$ or $J^2$ term plays the
central role (Bertin et al, 1989b; Bertin \& Lin 1996). The second
term of the RHS of equation (5.2) has been somehow ignored and
relegated to a residual $R$ term (see equation [B18] in Appendix B
of Lau \& Bertin 1978). Without this $J^2$ parameter, the cubic
dispersion relation simply reduces to the quadratic one in the
radial wavenumber $k$ for spiral density waves. In short, it is
this $T_1$ or $J^2$ term that is responsible for the third small
$k$ root of the so-called ``open mode" besides the familiar long-
and short-waves (see eq. [3.1] of Bertin et al. 1989b). In our
discussion here, this $J^2$ mechanism is referred to as the TSF
effect by its physical nature (LYF 2001).

The utmost reason we pinpoint this $B$ coefficient is as follows.
Bertin et al. (1989a, b) and Bertin \& Lin (1996) proposed the
cubic dispersion relation that was derived by keeping some
higher-order terms in the expansion of the Poisson equation. In
proper parameter regimes, two of the $k$ roots correspond to the
familiar short- and long-branches of spiral density waves while in
others, there exists a third small $k$ solution that was referred
to as the ``open mode" and was suggested to correspond to barred
spiral galaxies on the basis of their extensive numerical
calculations for the integro-differential equations of density
waves (Bertin et al. 1989a, b; Bertin \& Lin 1996). In contrast,
Shu et al. (2000) studied the two cases of aligned and unaligned
nonaxisymmetric stationary perturbations in isopedically
magnetized razor-thin SIDs and derived the marginal criteria for
both cases. Shu et al. (2000) suggested that the class of aligned
stationary perturbations is related to the secular barlike
instability in galactic dynamics.
This type of barlike ($m=2$) instability may be suppressed by
introducing a sufficiently massive dark-matter halo (Ostriker \&
Peebles 1973).
By numerically studying time-dependent overreflection process of
spiral waves across the corotation radius, Shu et al. (2000)
further recommended that the marginal criterion for unaligned
logarithmic stationary perturbations as an indicator for the onset
of spiral instabilities in the sense that they be eventually
amplified by a swing process near corotation (Goldreich \&
Lynden-Bell 1965; Mark 1976; Fan \& Lou 1997). By comparing their
marginal criterion (38) with their standard WKBJ dispersion
relation (39) for spiral density waves (Lin \& Shu 1966, 1968) and
by identifying an effective wavenumber
$(m^2+\alpha^2+1/4)^{1/2}r^{-1}$ as defined by their equation
(40), an analogy of quadratic ({\it rather than cubic}) form in
the effective wavenumber is apparent. Shu et al. (2000) proposed
to use their marginal criterion (38) in the limit of
$\alpha\rightarrow 0$ (i.e., ``breathing mode" limit; Lemos et al.
1991) for the onset of dynamical barred-spiral instability. The
question now is the connection or relation between the two
seemingly different proposals for the same physical problem in the
galactic context, namely the nature of {\it galactic bars and
barred spiral galaxies}.


The key is to realize the following basic fact. For the very SID
model (9) as prescribed by Shu et al. (2000), the third corotation
resonance term on the RHS of $B$ expression $(5.1)$
vanishes.\footnote{We could not identify the $B_s$ term attributed
to Lin \& Lau (1979) by Shu et al. (2000) in the paragraph
following their equation (53) (i.e. the last paragraph of their
subsection 5.1).
} For their SID model (9), the second term on the RHS of equation
$(5.1)$ becomes
$$
-{4m\Omega (d\nu/dr)\over\kappa r(1-\nu^2)}
=-{4m\Omega\omega\over\kappa^2r^2(1-\nu^2)}\ \eqno(5.3)
$$
which vanishes for $\omega=0$. Remarkably, a recombination of the
two terms on the RHS of equation (5.2) makes the $-T_1$ term or
$J^2$ term needed for the cubic dispersion relation (Bertin et al.
1989b) {\it disappear}. In reference to equation $(5.1)$ and the
stationarity requirement (i.e., $\omega=0$), it turns out that
$B=-m^2/r^2$.
Physically, this means that the TSF is absent for stationary
logarithmic spiral perturbations in the SID model (9) as given by
Shu et al. (2000) and should be the reason that in the analysis of
Shu et al. (2000), there is no obvious clue or counterpart for the
``open mode" as a root of a cubic dispersion relation. In fact,
equation (38) of Shu et al. (2000) together with their approximate
expression (36) for the Kalnajs function give a striking quadratic
form in terms of their effective wavenumber.

It seems plausible that the ``open mode" might correspond to the
``aligned" SID case (Shu 2000, private communications). It is then
the absence of the TSF for stationary logarithmic spiral
perturbations in SIDs that decouples the ``aligned" and
``unaligned" cases. In other words, the nonzero TSF for
time-dependent normal mode perturbations in differentially
rotating disks other than SIDs might somehow couple or mingle the
aligned and unaligned cases such that some kind of dispersion
relation in the spirit of a cubic dispersion relation emerges.
Although radial propagation is absent for aligned nonaxisymmetric
perturbations in a SID, there is an azimuthal wave propagation
relative to the axisymmetric background SID as made clear in our
analysis. Once present, the TSF might play a nontrivial coupling
role for nonaxisymmetric MSID perturbations. A carefully designed
numerical test may be needed to settle this issue.

We also note another line of reasoning. Even though equation (38)
of Shu et al. (2000) is approximately {\it quadratic} in terms of
their effective wavenumber $k=\alpha_e\tilde\omega^{-1}$ with
$|\alpha_e|\equiv (m^2+\alpha^2+1/4)^{1/2}$,
the recommended marginal criterion (37) (equivalently, eqs. [38]
or [41]) for unaligned stationary logarithmic spiral perturbations
yields only one solution of $\alpha$ for a given $D^2$ and $m\neq
0$ in a full SID (see Figure 3 of Shu et al. 2000). In other
words, by setting $D^2$
equal to a (sufficiently large, $D^2>0.5368$) constant in Figure 3
of Shu et al. (2000), there is only one intersection for a given
$m\neq 0$ and the corresponding root of $\alpha$ for a stationary
logarithmic spiral may be either large or small depending upon
whether $D^2$ is large or small. The point here is that the
situation of only one root of $\alpha$ for stationary logarithmic
spirals with $m\neq 0$ in a full SID does not necessarily
contradict the familiar result of short and long spiral density
waves (the two roots for the radial wavenumber $k$) for
propagating WKBJ spiral patterns in disks that are not full SIDs
(e.g., in partial SIDs).
By the same argument, the quadratic form of the marginal criterion
in terms of the effective wavenumber $\alpha_e\tilde\omega^{-1}$
for stationary logarithmic spirals in SIDs might not necessarily
contradict the cubic dispersion relation (Bertin et al. 1989;
Bertin \& Lin 1996) for nonstationary WKBJ-type spiral waves in
disks that are not full SIDs.


\subsection{Implications for the spiral galaxy NGC 6946}

The unambiguous case of interlaced optical and magnetic spiral
arms in the nearby spiral galaxy NGC 6946 was first reported by
Beck \& Hoernes (1996) in the almost rigidly rotating inner disk
portion. Similar interlaced arm features were also suspected
earlier in portions of spiral galaxies IC 342 (Krause et al. 1989)
and M83 (NGC 5236; Sukumar \& Allen 1989). These earlier
observations, the one of NGC 6946 in particular, prompted Fan \&
Lou (1996) to propose the concepts of FMDWs and SMDWs in
magnetized spiral galaxies (Lou \& Fan 1998a; LYF 2001).
Specifically, spiral perturbation enhancements of magnetic field
and gas mass density of SMDWs are significantly phase shifted
relative to each other (with a phase difference $\gsim\pi/2$). As
high-density gas arms are more vulnerable to cloud and star
formation activities and large-scale regular magnetic fields are
less disturbed by small-scale ISM turbulence associated with star
formation processes along phase-shifted magnetic arms, this
scenario naturally leads to interlaced optical and magnetic spiral
structures in magnetized disk galaxies. The recent wavelet
analysis on multi-wavelength data of NGC 6946 (Frick et al. 2000,
2001) revealed extended spiral arms well into the outer disk
portion with a flat rotation curve (e.g., Tacconi \& Young 1989;
Sofue 1996; Ferguson et al. 1998), which might appear to challenge
our earlier proposal that SMDWs be largely confined within the
inner disk portion of almost rigid rotation.

Given the idealizations in our MSID model, the two possibilities
of stationary logarithmic spiral fast and slow MHD perturbations
(see eq. [4.2.2]) in an MSID of flat rotation curve are
conceptually important for magnetized barred or normal spiral
galaxies in general (Lou \& Fan 2002) and may bear direct import
to the multi-wavelength data analysis on NGC 6946 by Frick et al.
(2000). In particular, for stationary slow MHD logarithmic spirals
in an MSID with flat rotation curve, spiral enhancements of
magnetic field and mass density are interlaced with a phase
difference $\gsim\pi/2$ by relation (4.2.1). For large $\alpha$ in
the WKBJ regime, this phase difference approaches $\sim\pi$, while
for small $\alpha\neq 0$ in the open regime, this phase difference
approaches $\sim\pi/2$. Our analytical solutions include the
effects of long-range self-gravity and disk differential rotation.
Therefore in magnetized spiral galaxies, slow MSID patterns can
indeed give rise to radially extended manifestations of interlaced
optical and magnetic spiral structures that persist well into the
outer disk portion with a largely flat rotation curve (Lou \& Fan
(2002).

We also note by equation (4.2.1) that for stationary fast
logarithmic spiral structures in a full or partial MSID with a
flat rotation curve, spiral enhancements of magnetic field and
mass density are in phase in the tight-winding regime but are
interlaced by a phase difference of $\sim\pi/2$ in the open
regime. This is a new feature that may bear consequences for
observations of galactic structures.

\subsection{Summary}

In reference to the work of Shu et al. (2000) on stationary
perturbation configurations of isopedically magnetized SID, we
have investigated both full and partial MSID stationary
perturbation configurations with a magnetic field coplanar with
the disk plane. Given the model specifications, we have reached
the following conclusions and suggestions.

(1) For the aligned case with $a^2>C_A^2/4$, there are two
possible values for the rotation parameter $D$ corresponding to
purely azimuthal propagations of FMDWs and SMDWs, respectively,
with distinctly different spatial phase relationships between
azimuthal magnetic field and surface mass density enhancements;
when $a^2<C_A^2/4$, there is only one valid value of $D^2$. Also,
the case of $m=0$ can be made to correspond to a rescaling of the
axisymmetric MSID background. Eccentric $|m|=1$ displacements may
occur for unconstrained $D^2$ values in a full MSID. In the
aligned case of barred configurations with $m=2$, there are two
different possible types of phase relationships between magnetic
field and mass density that should be worthwhile to search for
observationally.

(2) For a partial SID with a flat rotation curve, stationary
eccentric $|m|=1$ displacements are not allowed. For a partial
MSID with a flat rotation curve, stationary eccentric $|m|=1$
displacements can no longer occur for arbitrary $D^2$; they may
appear only when $a^2D^2=C_A^2/2$. For disk galaxies, it is the
usual case that $\Omega^2r^2=a^2D^2>C_A^2$ due to the presence of
massive dark-matter halos.
One needs other physical conditions to produce stationary
eccentric $|m|=1$ displacements in disk galaxies (e.g., power-law
rotation curves with $\beta\neq 0$; see Fig. 5 of Syer \& Tremaine
1996).

(3) For the unaligned logarithmic spiral case, there are two
values of $D^2$ for $|m|\ge 2$ corresponding to FMDWs and SMDWs
with distinct spatial pattern relationships between azimuthal
magnetic field and mass density enhancements. For $m=0$, there are
now two ring fragmentation regimes and one modified collapse
regime as shown in Fig. 1. In the absence of a coplanar magnetic
field, the small-$D^2$ ring fragmentation regime disappears.
In terms of phase relationships between azimuthal magnetic field
and surface mass density enhancements, the results in full and
partial MSIDs are basically the same. That is, in the
tight-winding regime, the two enhancements are in phase for fast
MSID configurations but are out of phase for slow MSID
configurations. For open structures, the two enhancements are
phase shifted by $\sim\pi/2$ for either fast and slow MSID
configurations.

(4) The MSID virial theorem (3.3.2) is suggestive that the
modified ratio ${\cal T}/|{\cal W}-{\cal M}|=D^2/[2(1+D^2)]$ be
crucial to determine the stability property of an MSID with a
coplanar azimuthal magnetic field, where ${\cal T}>0$ is the
rotational kinetic energy, ${\cal W}<0$ is the gravitational
potential energy, and ${\cal M}>0$ is the magnetic energy of the
entire MSID system.

(5) Regarding the conceptual connection between the cubic
dispersion relation in the modal formulation (Bertin et al. 1989a,
b; Bertin \& Lin 1996) and the perspective of stationary SID
configurations (Shu et al. 2000), we suggest that the absence of
the tangential shear force (TSF) in a SID of a flat rotation curve
decouples the bar modes (i.e., aligned configurations) from the
spiral modes (i.e., unaligned logarithmic spiral configurations).
This property is also carried over to our investigation of
stationary MSID configurations with a coplanar magnetic field.

(6) While our model formulation is idealized, the results provide
a conceptual basis and useful clues for diagnostics of galactic
bars and lopsided or barred and normal magnetized spiral galaxies.
For example, for stationary unaligned logarithmic spiral patterns
of SMDWs, the interlaced optical and radio-continuum spiral
structures may well extend into the disk domain with a largely
flat rotation curve as in the case of NGC 6946 (Lou \& Fan 2002).

\section*{Acknowledgments}

I thank F. H. Shu for valuable discussions. This research was
supported in part by grants from  US NSF (ATM-9320357 and
AST-9731623) to the University of Chicago, by the ASCI Center for
Astrophysical Thermonuclear Flashes at the University of Chicago
under Department of Energy contract B341495, by the Visiting
Scientist Program of the National Science Council
(NSC-88-2816-M-001-0010-6 and NSC89-2816-M001-0006-6) at the
Institute of Astronomy and Astrophysics, Academia Sinica, by the
Special Funds for Major State Basic Science Research Projects of
China, and by the Collaborative Research Fund from the National
Natural Science Foundation of China (NSFC) for Young Overseas
Chinese Scholars (NSFC 10028306) at the National Astronomical
Observatory, Chinese Academy of Sciences.
Institutional affiliations share this contribution.

\section{Appendix A}

The proof of determinant $\Delta>0$ in equation (3.2.10) goes as
follows. Before factorization (3.2.8) we solve equation (3.2.7)
with $F=1$
for two values of $a^2D^2$ as
\begin{eqnarray}
& & \!\!\!\!\!\!\!\! a^2D^2=
({\cal B}-{\cal A}-{\cal C}\pm\Delta^{1/2})
/[2(2-m^2-|m|)]\ ,\!\!\qquad\  (A1) \nonumber
\end{eqnarray}
where
$$
{\cal A}\equiv m^2 [a^2-a^2/|m|+C_A^2/(2|m|)] >0\ ,\eqno(A2)
$$
$$
{\cal B}\equiv 3C_A^2/2 >0\ ,\eqno(A3)
$$
$$
{\cal C}\equiv C_A^2( m^2-1/2 )( 1+1/|m| ) >0\ ,\eqno(A4)
$$
and the determinant $\Delta$
\begin{eqnarray}
& &
\Delta
=({\cal A}+{\cal B}-{\cal C})^2+C_A^4(m^2+|m|-2) \nonumber \\ &
&\qquad\qquad\qquad +2C_A^2(1-2/m^2){\cal A}
\ge 0\qquad\qquad\qquad\!\! (A5)\nonumber
\end{eqnarray}
for $|m|\ge 1$. In solution $(A1)$ for $a^2D^2$, one can factor
out $(|m|-1)$ based on the following three factorizations,
\begin{eqnarray}
& &\!\!\!\!\!\!\!\!\! [1-1/(2m^2)]{\cal A}-C_A^2/4
=(|m|-1)[(|m|+1){\cal A}/m^2
\nonumber \\ & &\qquad\qquad\qquad
+(a^2-C_A^2/2)/(2|m|)]\ , \qquad\qquad\quad\ (A6) \nonumber
\end{eqnarray}
$$
2(2-|m|^2-|m|)=-2(|m|-1)(|m|+2)\ ,\eqno(A7)
$$
\begin{eqnarray}
& &
{\cal A}-{\cal B}+{\cal C}=(|m|-1)[|m|a^2+C_A^2/2 \nonumber \\ &
&\qquad\qquad
+C_A^2(|m|+1)^2/|m|-C_A^2/(2|m|)]\ . \qquad\ \ (A8)\nonumber
\end{eqnarray}
This then completes the proof that the determinant $\Delta$ is
non-negative for $|m|\ge 1$ in solution (3.2.10) for $a^2D^2$.

%
%

\section{Appendix B}

As in Appendix A, we introduce handy notations
$$
{\cal A}\equiv m^2 [a^2-a^2F/|m|+C_A^2F/(2|m|)] >0\ ,\eqno(B1)
$$
$$
{\cal B}\equiv 3C_A^2/2 >0\ ,\eqno(B2)
$$
$$
{\cal C}\equiv C_A^2(m^2-1/2)(1+F/|m|) >0\ .\eqno(B3)
$$
The determinant $\Delta$ in solution (3.2.12) may be written as
\begin{eqnarray}
& &\!\!\!\!\!\!\Delta
=({\cal A}+{\cal B}-{\cal C})^2+C_A^4(m^2+F|m|-2) \nonumber \\ &
&\qquad\qquad +2C_A^2(1-2/m^2){\cal A}
> 0 \qquad\qquad\qquad\quad\ (B4) \nonumber
\end{eqnarray}
for $|m|\geq 1$. Both solutions of $a^2D^2$ in (3.2.12) are
positive.

\section{Appendix C}

Parallel to the proofs given in Appendices A and B, we outline the
proof here by introducing a set of handy notations
$$
{\cal A}\equiv [a^2-(a^2-C_A^2/2)F{\cal N}_m(\alpha)]
(m^2+\alpha^2+1/4)\ ,\eqno(C1)
$$
$$
{\cal B}\equiv C_A^2(m^2+\alpha^2-1/4)F{\cal N}_m(\alpha)\ ,
\eqno(C2)
$$
$$
{\cal C}\equiv C_A^2(m^2+\alpha^2-7/4)\ . \eqno(C3)
$$
The determinant $\Delta$ of equation (3.4.19) is given by
\begin{eqnarray}
& &\!\!\!\!\!\!\!\!\!\!\!\!\!\Delta\equiv ({\cal A}+{\cal B}+{\cal
C})^2 -4[m^2-2+(m^2+\alpha^2+1/4)F{\cal N}_m(\alpha)] \nonumber \\
& &\!\!\!\!\!\!\!\!\!\!\!\!\!\! \times [
C_A^2(m^2+\alpha^2-1/4){\cal A}/(m^2+\alpha^2+1/4)
-C_A^4/4 ] >0\ \!\ (C4) \nonumber
\end{eqnarray}
for $|m|\ge 2$. One needs to shuffle and regroup a few terms by
noting a $-4{\cal A}{\cal B}$ term and adding $2{\cal C}({\cal A}
-{\cal B})$ and $-2{\cal C}({\cal A}-{\cal B})$ terms and so
forth. It is useful to refer to Figure 1 of Shu et al. (2000) for
the magnitudes of ${\cal N}_m(\alpha)$ ($m=0, 1, 2$).

\section{Appendix D}

For aligned perturbation configurations constructed from a full
MSID of $F=1$, $D^2$ as given by equation (3.2.10) depends on
$q^2\equiv C_A^2/a^2$ and $|m|$. By solutions (3.2.10),
\begin{eqnarray}
& &\!\!\!\!\!\!\!\!\!\!\!\!
a^2D^2-C_A^2/2=[(|m|+1)C_A^2/2+a^2|m|^2]/[2|m|(2+|m|)] \nonumber
\\ & &\!\!\!\!\!\!\!\!\!\!\!\! \pm
[2|m|(2+|m|)]^{-1}\lbrace[C_A^2(|m|+1)/2+a^2|m|^2]^2 \qquad\quad\
\ \! (D1)\nonumber \\ & &\!\!\!\!\!\!\!\!\!\!\!\!
-2C_A^2|m|(2+|m|)(|m|+1)[a^2(|m|-1)-C_A^2|m|/2]\rbrace^{1/2}
\nonumber
\end{eqnarray}
%
An addition of $a^2$ to the first term on the RHS of solution
$(D1)$ gives
\begin{eqnarray}
& &\!\!\!\!\!\!\!\!\!\!\!\!\!
[(|m|+1)C_A^2/2+a^2|m|^2]/[2|m|(2+|m|)]+a^2 \nonumber \\ &
&\!\!\!\!\!\!\!\!\!\!\!\!\!
=[(|m|+1)C_A^2/2+3a^2|m|^2+4a^2|m|]/[2|m|(2+|m|)]\ . \
(D2)\nonumber
\end{eqnarray}
Examine then the determinant $\Delta$ in solutions (3.2.10) or
$(D1)$ in the form of
\begin{eqnarray}
& &\!\!\!\!\!\!\!\!\!\!\!\!\! [(|m|+1)C_A^2/2+3a^2|m|^2+4a^2|m|]^2
\nonumber \\ & &\!\!\!\!\!\!\!\!\!\!\!\!\!
-4a^2(m^2+2|m|)(|m|+1)(2a^2|m|+C_A^2/2) \nonumber \\ &
&\!\!\!\!\!\!\!\!\!\!\!\!\!
-2C_A^2|m|(2+|m|)(|m|+1)[a^2(|m|-1)-C_A^2|m|/2]\ . \ \ \ \! (D3)
\nonumber
\end{eqnarray}
The sum of the last two terms of expression $(D3)$ is
$$
-8(|m|^2+2|m|)(|m|+1)|m|(a^2-C_A^2/4)(a^2+C_A^2/2)\ .\eqno(D4)
$$
When $a^2>C_A^2/4$, the background surface mass density
$\Sigma_{0}>0$ for both the plus- and minus-sign solutions
(3.2.10) of $a^2D^2$. For $a^2<C_A^2/4$, one has $\Sigma_0>0$ only
for the plus-sign solution, while the minus-sign solution gives
$\Sigma_0<0$.

For the numerator $2\Omega^2r^2-C_A^2/2=2(a^2D^2-C_A^2/2+C_A^2/4)$
on the RHS of equation (3.1.9), we add $C_A^2/4$ to the first term
on the RHS of solution $(D1)$ to derive
\begin{eqnarray}
& & {(|m|+1)C_A^2/2+a^2|m|^2\over 2|m|(2+|m|)}+{C_A^2\over 4}
\nonumber \\ & &\qquad ={(|m|^2+3|m|+1)C_A^2/2+a^2|m|^2\over
2|m|(2+|m|)}>0\ . \qquad (D5) \nonumber
\end{eqnarray}
The determinant $\Delta$ may be rearranged into the form of
\begin{eqnarray}
& &\!\!\!\!\!\! \Delta
=[(|m|^2+3|m|+1)C_A^2/2+a^2|m|^2]^2 \nonumber \\ & & \quad
-C_A^2(|m|^2+2|m|)(3|m|^2-2)(a^2-C_A^2/4)\ . \qquad\!  (D6)
\nonumber
\end{eqnarray}
For $a^2-C_A^2/4>0$ in equation $(D6)$, one has both
$\Sigma_{0}>0$ and $2a^2D^2-C_A^2/2>0$ for either plus- or
minus-solutions in equation (3.2.10). For $a^2-C_A^2/4<0$, only
the plus-sign solution with $a^2D^2-C_A^2/4>0$ is physically
valid.

The denominator on the RHS of equation (3.1.9) is
$$
\!\! m^2\Omega^2r^2-C_A^2(m^2-1/2)=
m^2[\Omega^2r^2-C_A^2+C_A^2/(2m^2)]\ .\eqno(D7)
$$
A subtraction of $C_A^2-C_A^2/(2m^2)$ from the first term on the
RHS of solution (3.2.10) gives
\begin{eqnarray}
& &\!\!\!\!\!\!\!\!\!\!\!\! {C_A^2/2+5C_A^2|m|/2+(a^2+C_A^2)|m|^2
\over 2|m|(2+|m|)}-\bigg(1-{1\over 2m^2}\bigg)C_A^2 \nonumber \\ &
&\!\!\!\!\!\!\!\! ={3C_A^2(1-|m|)/2+(a^2-C_A^2)|m|^2+2C_A^2/|m|
\over 2|m|(2+|m|)}\ .\qquad (D8) \nonumber
\end{eqnarray}
The determinant $\Delta$ is now examined in the form of
\begin{eqnarray}
& &\!\!\!\!\!\!\!\!\! \Delta
=[3C_A^2(1-|m|)/2+(a^2-C_A^2)|m|^2+2C_A^2/|m|]^2 \nonumber \\ & &
\ +C_A^4(|m|+2)(|m|+1)(3|m|^2-2)/|m|^2 >0\ \qquad (D9) \nonumber
\end{eqnarray}
for $m\neq 0$.

For the sign of the background surface mass density $\Sigma_{0}$
of a partial MSID with $|m|\ge 1$, we consider
$a^2D^2+a^2-C_A^2/2$. By adding $a^2-C_A^2/2$ to the term
containing ${\cal B}-{\cal A}-{\cal C}$ on the RHS of solution
(3.2.12) for $a^2D^2$, one has
$$
{m^2a^2+2a^2(m^2+F|m|/2-2)+(|m|-|m|^{-1})C_A^2F/2 \over
2(m^2+F|m|-2)}>0\ .\eqno(D10)
$$
In the determinant $\Delta$ of solution (3.2.12), we then consider
the following part
\begin{eqnarray}
& &\!\!\!\!\!\!\!\!\!\! -2(a^2-C_A^2/2)(m^2+F|m|-2)
\qquad\qquad\qquad\qquad\quad (D11)\nonumber \\ & &\quad
\!\!\!\!\!\!\!\!\!\!\!\!\!\!\!\! \times
[2(a^2-C_A^2/2)(m^2+F|m|-2)
+2( {\cal A}-{\cal B}+{\cal C})]
\nonumber \\ & &\!\!\!\!\!\!\!\!\!\! +4C_A^2[(m^2-1/2){\cal A}/m^2
-C_A^2/4](2-m^2-F|m|) \nonumber \\ & &\!\!\!\!\!\!\!\!\!
=-8(m^2+F|m|-2)(m^2-1)
(a^2+C_A^2/2)(a^2-C_A^2/4)\ ,
\nonumber
\end{eqnarray}
where ${\cal A}$, ${\cal B}$, ${\cal C}$ are defined by equations
$(B1)$, $(B2)$, $(B3)$ in Appendix B. When $a^2-C_A^2/4>0$,
$\Sigma_0$ is positive for solutions (3.2.12) of both signs; for
$a^2-C_A^2/4<0$, $\Sigma_0$ is positive for the plus-sign
solution, while $\Sigma_0<0$ for the minus-sign solution.

For the numerator on the RHS of (3.1.9), we consider
$a^2D^2-C_A^2/4$ by subtracting $C_A^2/4$ from the term involving
${\cal B}-{\cal A}-{\cal C}$ on the RHS of solution (3.2.12) and
obtain
$$
{a^2(m^2-|m|F)+(m^2+2|m|F-2-F/|m|)C_A^2/2 \over 2(m^2+F|m|-2)}>0\
.\eqno(D12)
$$
For the determinant $\Delta$ of solution (3.2.12), we then
consider the following part
\begin{eqnarray}
& & \!\!\!\!\!\!\!\!\!\!\!\!\! C_A^2(m^2+F|m|-2) [{\cal A}-{\cal
B}+{\cal C}
-C_A^2(m^2+F|m|-2)/4]
\nonumber \\ & &\!\!\!\!\!\!\!\!\!\!\!\!\! +4C_A^2 [(m^2-1/2){\cal
A}/m^2
-C_A^2/4 ](2-m^2-F|m|) \qquad (D13)\nonumber \\ &
&\!\!\!\!\!\!\!\!\!\!\!\!\! =-C_A^2(m^2+F|m|-2)
(3m^2-2)(a^2-C_A^2/4)(1-F/|m|),
\nonumber
\end{eqnarray}
where ${\cal A}$, ${\cal B}$, ${\cal C}$ are defined
in Appendix B.

For the sign of the denominator on the RHS of equation (3.1.9), we
consider $a^2D^2-[1-1/(2m^2)]C_A^2$. In the determinant $\Delta$
of solution (3.2.12), we consider the portion
\begin{eqnarray}
& & \!\!\!\!\!\!\!\!\! (2-1/m^2)C_A^2(m^2+F|m|-2) \nonumber \\ &
&\quad \!\!\!\!\!\!\!\!\!\!\!\! \times [2({\cal A}-{\cal B}+{\cal
C})
-(2-1/m^2)C_A^2 (m^2+F|m|-2)] \nonumber \\ &
&\qquad\!\!\!\!\!\!\!\!\!\!\!\! +4C_A^2 [(m^2-1/2){\cal A}/m^2
-C_A^2/4](2-m^2-F|m|) \nonumber \\ & &\!\!\!\!\!\!\!\!
=C_A^4(m^2+F|m|-2)(m^2-1)(3m^2-2)/m^4\ . \qquad\ (D14)\nonumber
\end{eqnarray}
Thus, $m^2\Omega^2r^2-C_A^2(m^2-1/2)$ is positive and negative for
the plus- and minus-sign solutions of (3.2.12), respectively.

\section{Appendix E}

We consider the following factor that appears in criterion
(3.4.16) for logarithmic spirals in an MSID,
\begin{eqnarray}
& &\!\!\!\!\!\!\!\!\!\!\!\!\!\!
(m^2+\alpha^2+1/4)\Omega^2r^2-(m^2+\alpha^2-1/4)C_A^2
\qquad\qquad\ \ \  (E1) \nonumber \\ &
&\!\!\!\!\!\!\!\!\!\!\!\!\!\!
=(m^2+\alpha^2+1/4)\bigg\lbrace\Omega^2r^2 -\bigg[1-{1\over
2(m^2+\alpha^2+1/4)}\bigg]C_A^2\bigg\rbrace\ . \nonumber
\end{eqnarray}
Using the stationary criterion in the form of equation (3.4.19),
we derive two values of $a^2D^2$
$$
a^2D^2={{\cal A}+{\cal B}+{\cal C}\pm\Delta^{1/2} \over
2[m^2-2+(m^2+\alpha^2+1/4)F{\cal N}_m(\alpha)]}\ ,\eqno(E2)
$$
where ${\cal A}$, ${\cal B}$, ${\cal C}$, and $\Delta$ are defined
by equations $(C1)-(C4)$ in Appendix C.
The determinant $\Delta$
is non-negative for $|m|\geq 2$.
We subtract $C_A^2\{1-[2(m^2+\alpha^2+1/4)]^{-1}\}$ from solution
$(E2)$ and rearrange the determinant $\Delta$ of solution $(E2)$.
In the determinant $\Delta$, we consider the following part,
\begin{eqnarray}
& &\!\!\!\!\!\!\!\!\!\!\! [m^2-2+(m^2+\alpha^2+1/4)F{\cal
N}_m(\alpha)] \nonumber \\ & &\!\!\!\!\!\!\!\!\! \times
[2-1/(m^2+\alpha^2+1/4)]C_A^2 \nonumber \\ & &\!\!\!\!\!\!\!\!\!
\times\lbrace 2({\cal A}+{\cal B}+{\cal C})
-[m^2-2+(m^2+\alpha^2+1/4)F{\cal N}_m(\alpha)] \nonumber \\ &
&\!\!\!\!\!\!\!\!\! \times[2-1/(m^2+\alpha^2+1/4)]C_A^2\rbrace
\nonumber \\ & &\!\!\!\!\!\!\!\!\!
-4[m^2-2+(m^2+\alpha^2+1/4)F{\cal N}_m(\alpha)] \nonumber \\ &
&\!\!\!\!\!\!\!\!\! \times [ C_A^2(m^2+\alpha^2-1/4){\cal A}
/(m^2+\alpha^2+1/4)
-C_A^4/4 ] \nonumber \\ & & =C_A^4 [m^2-2+(m^2+\alpha^2+1/4)F{\cal
N}_m(\alpha)] \nonumber \\ & &\quad \times\lbrace
[2-1/(m^2+\alpha^2+1/4)][2(\alpha^2+1/4) \nonumber \\ & &\qquad
+(m^2-2)/(m^2+\alpha^2+1/4)]+1\rbrace\ , \qquad\qquad\ (E3)
\nonumber
\end{eqnarray}
which is positive for $|m|\ge 2$.

\begin{figure}
  \vspace{50pt}
  \caption{ Parameter regimes in terms of $D^2$, $\alpha$, and
            $q^2\equiv C_A^2/a^2$ separated by stationary MSID
            configurations for unaligned logarithmic spirals with
            $m=0$. (a) $q^2=3.61$; (b) $q^2=1.0$; (c) $q^2=0.09$. }
\end{figure}


\clearpage
\begin{figure}
\begin{center}
\includegraphics[scale=1]{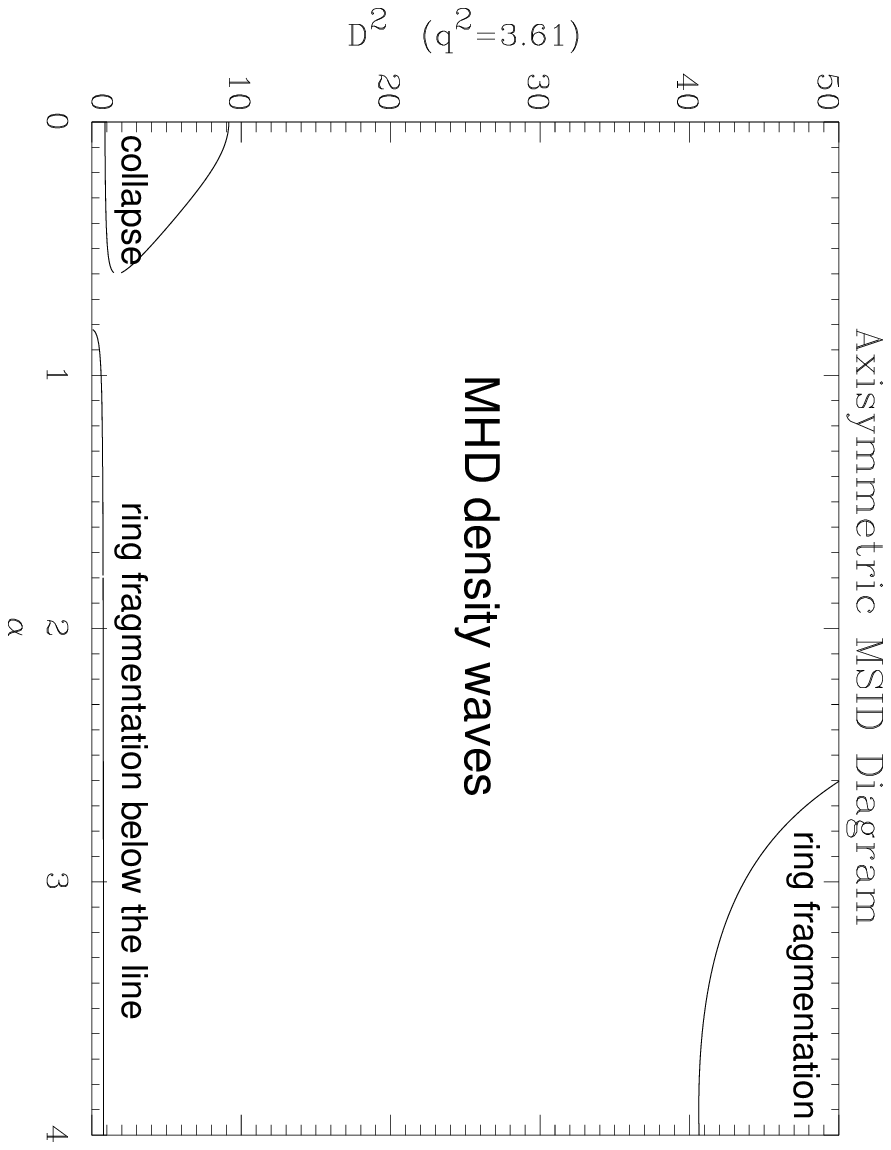}
\end{center}
\end{figure}






\clearpage
\begin{figure}
\begin{center}
\includegraphics[scale=1]{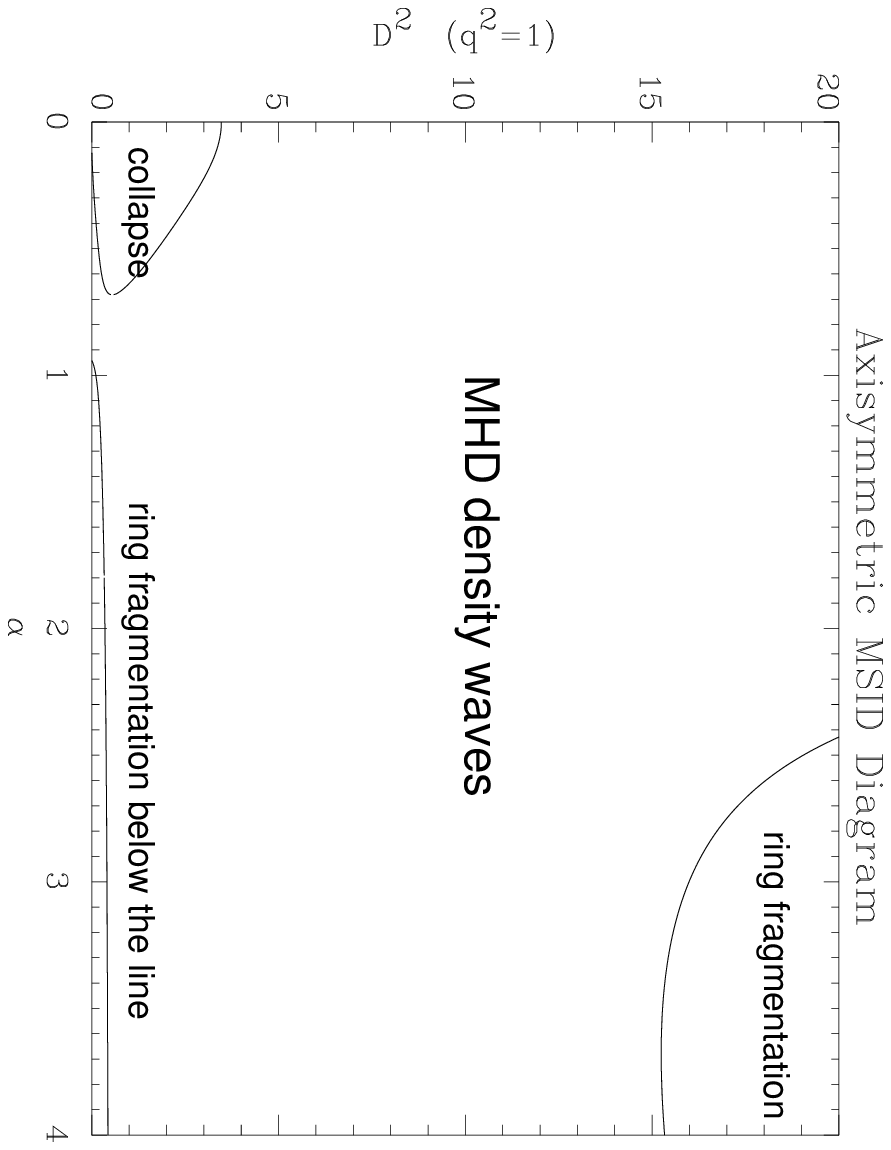}
\end{center}
\end{figure}


\clearpage
\begin{figure}
\begin{center}
\includegraphics[scale=1]{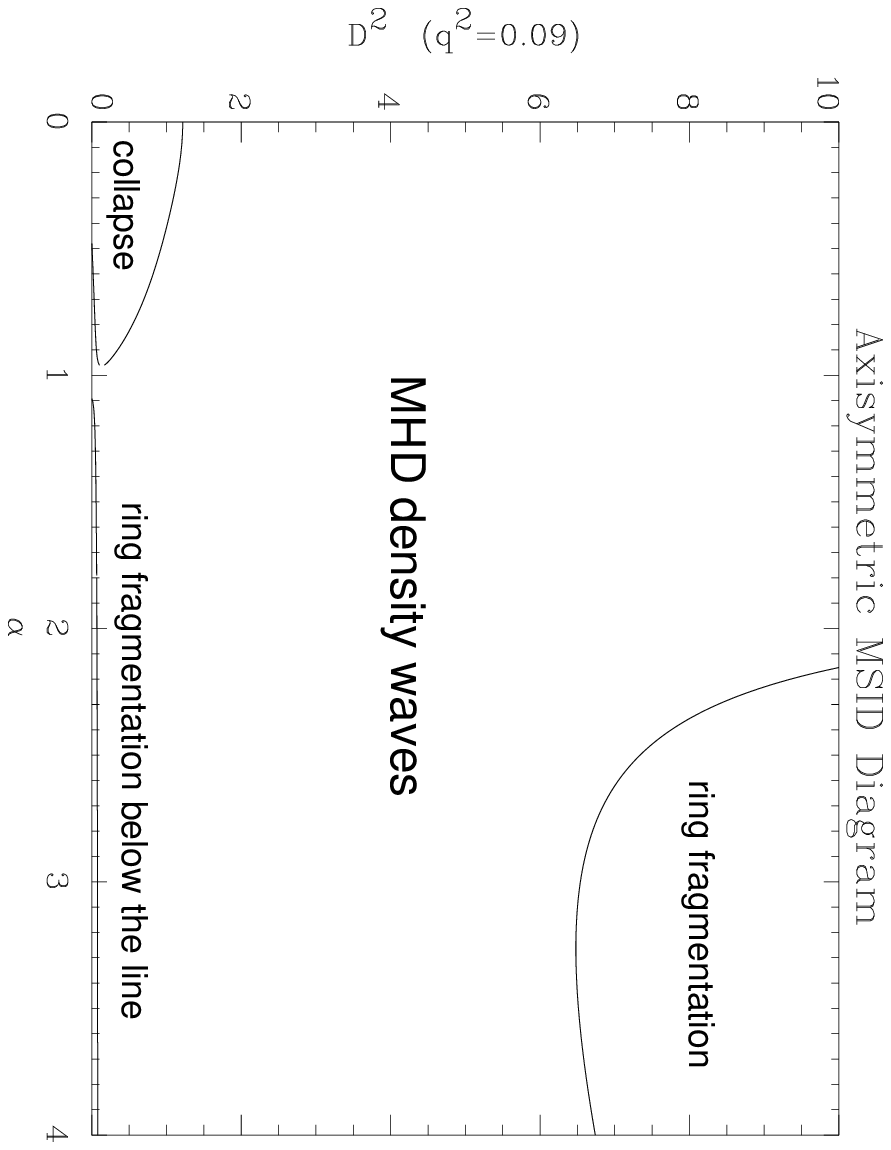}
\end{center}
\end{figure}

\label{lastpage}

\end{document}
===================
\documentclass[useAMS]{mn2e}
\usepackage{graphicx}
\begin{document}
\includegraphics{q2l.eps}
\end{document}